\documentclass[12pt,titlepage]{article}
\usepackage{graphicx}
\usepackage{natbib}

\usepackage[left=2cm, right=2cm, top=2cm]{geometry}

\usepackage[]{hyperref}
\hypersetup{colorlinks=true,citecolor=blue}

\usepackage{amsmath}
\usepackage{amssymb}
\usepackage{txfonts}
\usepackage{color}

\usepackage{authblk}

\title{The environmental effects of very large bolide impacts on early Mars explored with a hierarchy of numerical models}
\author[1]{Martin Turbet}
\author[2,3]{Cedric Gillmann}
\author[1]{Francois Forget}
\author[1,4]{Baptiste Baudin}
\author[5]{Ashley Palumbo}
\author[5]{James Head}
\author[2]{Ozgur Karatekin}

\affil[1]{Laboratoire de M\'et\'eorologie Dynamique/IPSL, CNRS, Sorbonne Universit\'e, Ecole normale sup\'erieure, PSL Research University, Ecole Polytechnique, 75005 Paris, France.}
\affil[2]{Royal Observatory of Belgium, Brussels, Belgium}
\affil[3]{Free University of Brussels, Department of Geosciences, G-Time, Brussels, Belgium}
\affil[4]{Magist\`ere de Physique Fondamentale, D\'epartement de Physique, Univ. Paris-Sud, Universit\'e Paris-Saclay, 91405 Orsay Campus, France.}
\affil[5]{Department of Earth, Environmental, and Planetary Sciences, Brown University, Providence, RI02912, USA.}

\date{\today}

\begin{document}

\maketitle

\begin{abstract}

The origin of the presence of geological and mineralogical evidence that liquid water 
flowed on the surface of early Mars is now a 50-year-old mystery.
It has been proposed \citep{Segura:2002,Segura:2008,Segura:2012} that bolide impacts could have triggered a long-term climate change, 
producing precipitation and runoff that may have altered the surface of Mars in a way that could explain (at least part of) this evidence.
Here we use a hierarchy of numerical models (a 3-D Global Climate Model, a 1-D radiative-convective model and a 
2-D Mantle Dynamics model) to test that hypothesis and more generally explore the 
environmental effects of very large bolide impacts (D$_{\text{impactor}}$~$>$~100~km, 
or D$_{\text{crater}}$~$>$~600~km) on the atmosphere, 
surface and interior of early Mars. 

Using a combination of 1-D and 3-D climate simulations, 
we show that the environmental effects of the largest impact events recorded on Mars are characterized by: (i) 
a short impact-induced warm period (several tens of Earth years for the surface and atmosphere to be back to ambient conditions after 
very large impact events); (ii) a low amount of hydrological cycling of water 
(because the evaporation of precipitation that reached the ground is extremely limited). 
The total cumulative amount of precipitation (rainfall) can be reasonably well approximated by the initial post-impact atmospheric 
reservoir of water vapour (coming from the impactor, 
the impacted terrain and from the sublimation of permanent ice reservoirs 
heated by the hot ejecta layer); 
(iii) deluge-style precipitation ($\sim$2.6~m Global Equivalent Layer of surface precipitation per Earth year for 
our reference simulation, 
quantitatively in agreement with previous 1-D cloud free climate calculations of \citealt{Segura:2002}), 
and (iv) precipitation patterns that are uncorrelated with the observed regions of valley networks. 

However, we show that the impact-induced 
stable runaway greenhouse state predicted by \citet{Segura:2012} 
should not be achievable if convection and water vapour condensation processes are considered. 
We nevertheless confirm the results of \citet{Segura:2008} and \citet{Urata:2013h2o}
that water ice clouds could in theory significantly 
extend the duration of the post-impact warm period, and even for cloud coverage
significantly lower than predicted in \citet{Ramirez:2017}. However, 
the range of cloud microphysical properties for which this scenario works is very narrow. 

Using 2-D Mantle Dynamics simulations  we find that large bolide impacts 
can produce a strong thermal anomaly in the mantle of Mars that can survive 
and propagate for tens of millions of years. 
This thermal anomaly could raise the near-surface internal heat flux up to several hundreds 
of mW/m$^2$ (i.e. up to $\sim$~10~times the ambient flux) for several millions years at the edges of the impact crater. 
However, such internal heat flux is largely insufficient to keep the martian surface above the melting point of water.

In addition to the 
poor temporal correlation between the formation of the largest basins and valley networks \citep{Fassett:2011}, 
these arguments indicate that the largest 
impact events are unlikely to be the direct cause of formation of 
the Noachian valley networks. Our numerical results support instead the prediction of \citet{Palumbo:2018impact} 
that very large impact-induced rainfall could have caused degradation of large craters, 
erased small craters, and formed smooth plains, potentially erasing much of the previously visible morphological surface history. 
Such hot rainfalls may have also led to the formation of aqueous alteration products on 
Noachian-aged terrains, which is consistent with the timing of formation of clays.

\end{abstract}


\section{Introduction}

Since the 1970s, scientists have discovered an overwhelming number of pieces of evidence that 
liquid water flowed on ancient Mars: high erosion rates \citep{Craddock:2002,Mangold:2012,Quantin:2019}, 
sedimentary deposits \citep{Williams:2013,Grotzinger:2015}, 
hydrated minerals \citep{Poulet:2005,Bibring:2006,Mustard:2008,Ehlmann:2011,Carter:2013,Carter:2015} and dry 
river beds and lakes \citep{Carr:1995,Cabrol:1999,Malin:2003,Moore:2003,Mangold:2006,Hynek:2010}. 
Sophisticated climate modelling under ancient Mars conditions assuming a faint young Sun and CO$_2$/N$_2$/H$_2$O atmospheres 
have not yet been able to produce liquid water or significant precipitation anywhere on the 
red planet \citep{Forget:2013,Wordsworth:2013}, unless additional hypothetic greenhouse gases 
were incorporated, such as reducing gases CH$_4$ or H$_2$ 
\citep{Ramirez:2014,Wordsworth:2017,Kite:2017,Ramirez:2017b,Ramirez:2018,Turbet:2019icarus}.

It has been suggested that the warmer and wetter (compared to present-day Mars) conditions 
required to explain the formation of the aforementioned geological and mineralogical evidence 
could have been produced in response to 
impact-induced climate change \citep{Segura:2002,Segura:2008,Toon:2010,Wordsworth:2013,Wordsworth:2016,Haberle:2017,turbet_phd:18,Steakley:2018}. 


The environmental effects of such bolide impact events have already been studied with unidimensional 
radiative-convective models \citep{Segura:2002,Segura:2008,Segura:2012}. 
Here we improve upon this previous work by exploring the climatic impact 
of very large bolide impact events using a hierarchy of numerical models, ranging from 
(i) a 3-D Global Climate Model to (ii) a 1-D radiative-convective model and ultimately to (iii) a 2-D mantle dynamics
numerical code. The 3-D Global Climate Model (GCM) simulations 
are used to accurately simulate all the processes (cloud microphysics, large scale circulation, etc.) at 
play in the post-impact early martian atmosphere. 1-D radiative-convective model simulations are used to simulate 
a wide range of possible post-impact conditions (initial atmosphere, size of impactor, etc.) and explore 
the sensitivity to parameterizations (e.g. microphysics of clouds), thanks to their low computational cost.
Eventually, the 2-D mantle dynamics numerical code is used to explore the long-term effects of large bolide impacts 
on the surface and near-surface environment on ancient Mars.

This paper focuses on the environmental effects of the largest impact events ever recorded on Mars, i.e. 
those that are large enough (impactor diameter typically larger than 100~km, 
corresponding to a crater diameter roughly larger than 600~km, using \citealt{Toon:2010} scaling relationship) 
to typically vaporize the equivalent of the present-day Mars water content (around 34~m Global Equivalent Layer [GEL]; \citealt{Carr:2015}) in 
the atmosphere, as estimated from energy conservation calculations. 
Although the formation of the largest basins visible today on Mars (e.g. Hellas, Argyre, Isidis) is now thought to 
have occured earlier in time than the formation of the martian valley networks \citep{Fassett:2008,Fassett:2011,Werner:2014} 
(compared to \citealt{Toon:2010} that used basin age estimates of \citealt{Werner:2008}), 
very large impact events must have had a profound impact on the surface of early Mars.

The specific processes that occur following a basin-scale impact event on Mars have been explored by 
\citet{Segura:2002,Segura:2008} and \citet{Toon:2010} and revisited from a geological perspective by \citet{Palumbo:2018impact}. 
Before discussing the specific modeling done in this work, 
we first re-iterate the key processes involved in Impact Cratering Atmospheric 
and Surface Effects (ICASE) \citep{Palumbo:2018impact}. 
When a very large impactor collides with the martian surface, 
a significant amount of energy is transferred from the impactor to the surface. 
The intense energy of the impact event will cause all projectile material and some target martian material 
to be pulverized (and ejected from the crater), melted (forming a melt layer on the crater floor), 
or vaporized. The vaporized component is of specific interest for this analysis. 
The vaporized material expands and moves away from the crater, producing an extremely hot 
plume consisting of target and projectile material. 
The specific constituents of the plume include water vapour and vaporized silicate material. 
For large, energetic impact events such as the ones explored in the present work, 
the plume will expand globally. 
Atmospheric temperatures are expected to be very hot, above the condensation temperature of 
both the silicate material and water. 
The atmosphere begins cooling from the initial extremely hot state and, due to the differences in condensation temperature, 
the silicate material is expected to condense and fall out of the atmosphere before the water vapour. 
Upon condensation, the silicate material is expected to fall out of the atmosphere and distribute globally, 
forming something similar to a very hot terrestrial spherule layer. 
The high temperature of this silicate-debris layer will cause any remaining underlying water ice to 
vaporize and enter the atmosphere. At this point, we expect that the entire initial surface water 
inventory would be as vapour in the atmosphere. 
Based on this description of the post-impact effects that occur following a basin-scale impact event on Mars 
(following \citealt{Palumbo:2018impact}), we set out to constrain the duration and characteristics of 
impact-induced rainfall using a hierarchy of numerical models. 
Specifically, we explore two main aspects of post-impact effects in this manuscript:
\begin{enumerate}
    \item How long can the surface of Mars be kept above the melting point of water following 
large bolide impact events? In particular, can an impact-induced, stable runaway 
climate exist on early Mars, as previously reported by \citet{Segura:2012}? 
Can impact-induced, high altitude water ice clouds keep the surface of Mars above 
the melting point of water for extended periods of time, as previously reported 
by \citet{Segura:2008} and \citet{Urata:2013h2o}?
    \item How much precipitation is generated after large bolide impacts events, and how is it distributed across the surface of Mars?
\end{enumerate}
A major, original aspect of our work is to explore how 3-dimensional processes 
(atmospheric circulation and cloud formation) affect the environmental effects of bolide impact events.

We first describe the various numerical models used in this work in Section~\ref{impact_method}: 
(i) The 3-D LMD Generic Global Climate Model, (ii) the 1-D LMD Generic inverse radiative-convective model and 
(iii) the StagYY 2-D mantle dynamics code. We then present in Section~\ref{impact_3D} the results of 
our 3-D global climate simulations of the environmental effects of large bolide impact events on the atmosphere and 
surface of early Mars. 
Because we show that in some conditions 3-D post-impact atmospheres can be remarkably well described by 1-D simulations, 
we then use our 1-D radiative-convective simulations in Section~\ref{impact_1D} to explore 
a wide range of possible impact-induced conditions. Eventually,
we use in Section~\ref{stagyy_part} our 2-D mantle dynamics numerical simulations to model the long-term effects of very 
large impacts on the interior, near-surface and surface of Mars. We revisit throughout the manuscript the results 
of \citet{Segura:2002,Segura:2008,Segura:2012,Urata:2013h2o} and \citet{Ramirez:2017}.

\section{Method}
\label{impact_method}

In this section, we describe the hierarchy of numerical models used for the present study.
We first describe the two different versions (3-D and 1-D) of the LMD Generic climate model, designed here 
to reproduce the post-impact conditions following a very large impactor hitting the surface of early Mars. 
We assume that the planet - initially endowed with a CO$_2$-dominated atmosphere - is suddenly warmed and moistened 
following processes described in \citet{Palumbo:2018impact} and summarized in the previous section. 
We then present the radiative transfer scheme (common for the two models), 
with a particular focus on the recent improvements made on 
the spectroscopy of dense CO$_2$+H$_2$O atmospheres (typical of post-impact atmospheres) that are taken into account in 
the radiative transfer calculations. Eventually, we describe the StagYY 2-D mantle dynamics code used to 
model the long-term effects of very large impacts on the surface and near-surface of Mars.


\subsection{3-D Global Climate Model simulations}
\label{large_impact_3Dmodel_method}

The model described in this subsection was used to produce the results described in Section~\ref{impact_3D}.

Our 3-D LMD Generic model is a full 3-Dimensions Global Climate Model (GCM) that has previously been developed and 
used for the study of the climate of ancient Mars \citep{Forget:2013,Wordsworth:2013,Wordsworth:2015,Turbet:2017icarus,Palumbo:2018}.

The simulations presented in this paper were performed at a spatial resolution of 96~x~64 in 
longitude~x~latitude (i.e. 3.8$^{\circ}$~$\times$~2.8$^{\circ}$; 220~km~x~165~km at the equator). In the vertical direction, the model is composed of 45~distinct atmospheric layers, ranging from the surface up to a few Pascals. Hybrid $\sigma$ coordinates (where $\sigma$ is the ratio between pressure and surface pressure) and fixed pressure levels were used in the lower and the upper atmosphere, respectively. 

The dynamical time step of the simulations ranged between 9~s (at the beginning of the large impact events) and 90~s. The radiative transfer (described in subsection~\ref{impact_rad_transfer}) and the physical parameterizations (such as condensation, convection, etc.) are calculated every 10 and 40 dynamical time steps, respectively.

Recent work has suggested that the Tharsis rise may have been largely emplaced after the formation of the valley 
networks (e.g. \citealt{Bouley:2016}). Thus, at the time of the large basin impact events, which is now thought to have occurred 
earlier in time than valley network formation \citep{Fassett:2008,Fassett:2011,Werner:2014}, we assume topography that 
is consistent with pre-Tharsis conditions. Specifically, we used the pre-True Polar Wander (pre-TPW) topography from \citet{Bouley:2016}. 
The pre-TPW topography is based on the present-day MOLA (Mars Orbiter Laser Altimeter) Mars surface topography 
\citep{Smith:1999,Smith:2001}, but without Tharsis and all the younger volcanic features. Moreover, the formation 
of Tharsis should have produced a large True Polar Wander (TPW) event of 20$^{\circ}$-25$^{\circ}$, which is also 
taken into account in the pre-TPW topography.

We set the obliquity of Mars at 40$^{\circ}$ to be roughly consistent with the most statistically likely obliquity (41.8$^{\circ}$) for 
ancient Mars \citep{Laskar:2004}. We also set the eccentricity to zero as in \citet{Wordsworth:2013}.


To account for the thermal conduction in the subsurface, we used a 19-layers thermal diffusion soil model. The mid-layer depths 
range from d$_0$~$\sim$~0.15~mm to d$_{19}$~$\sim$~80~m, following the power law 
d$_n$~=~d$_0$~$\times$~2$^n$ with $n$ being the corresponding soil level, chosen to take 
into account both the diurnal and seasonal thermal waves.
We assumed the thermal inertia of the regolith I$_{\text{ground}}$ to be equal to:
\begin{equation}
    I_{\text{ground}}=I_{\text{dry}}+7~x_{\text{H}_2\text{O}}, 
\end{equation}
where I$_\text{dry}$~=~250~J~m$^{-2}$~s$^{-1/2}$~K$^{-1}$ and x$_{\text{H}_2\text{O}}$ is the 
soil moisture (in kg~m$^{-3}$). 
The soil moisture is calculated in the first meter of the ground only. 
More information on this parameterization can be found in \citealt{turbet_phd:18} (Chapter~9 and Figure~9.1).
The dry regolith thermal inertia is slightly higher than the present-day 
Mars global mean thermal inertia in order to account for the higher atmospheric pressure \citep{Piqueux:2009}. 
This expression has been derived from the standard parameterization of the ORCHIDEE (Organising Carbon and Hydrology 
In Dynamic Ecosystems) Earth land model \citep{Wang:2016}.
Moreover, we arbitrarily fixed the thermal inertia of the ground to a value 
of 1500~J~m$^{-2}$~s$^{-1/2}$~K$^{-1}$, whenever the snow/ice cover exceeds a threshold of 1000~kg~m$^{-2}$ (i.e. the snow/ice cover thickness locally exceeds 1~m).
We assumed that the martian regolith has a maximum water capacity of 150~kg~m$^{-2}$, based on a 
simple bucket model widely used in the Earth land 
community \citep{Manabe:1969,Wood:1992,Schaake:1996}.

Subgrid-scale dynamical processes (turbulent mixing and convection) were parameterized as in \citet{Forget:2013} and \citet{Wordsworth:2013}. The planetary boundary layer was accounted for by the \citet{Mellor:1982} and \citet{Galperin:1988} time-dependent 2.5-level closure scheme, and complemented by a convective adjustment which rapidly mixes the atmosphere in the case of unstable temperature profiles.
Moist convection was taken into account following a moist convective adjustment that originally derives from the 'Manabe scheme' \citep{Manabe:1967,Wordsworth:2013}. In the version of our scheme, relative humidity is let free and limited to 100$\%$ (supersaturation is not permitted). This scheme was chosen instead of more refined ones because it is: 1. robust for a wide range of pressures; 2. energy-conservative; and 3. it is a physically consistent scheme for exotic (non Earth-like) situations such as the ones induced by large bolide impact events. In practice, when an atmospheric grid cell reaches 100~$\%$ saturation and the corresponding atmospheric column has an unstable temperature vertical profile, the moist convective adjustment scheme is performed to get a stable moist adiabatic lapse rate. 
In our simulations of large impact events, water vapour can become the dominant atmospheric species. Thus, we used a generalized formulation of the moist-adiabiat lapse rate developed by \citet{Leconte:2013nat} (Supplementary Materials) to account for the fact that water vapour can become a main species in our simulations. In our model we also used the numerical scheme proposed by \citet{Leconte:2013nat} (Supplementary Materials) to account for atmospheric mass change after the condensation or the evaporation of gases (water vapour in our case); this numerical scheme is crucial in our simulations of impact events to model accurately the evolution of the surface pressure and the relative content of CO$_2$ and H$_2$O. More details on the scheme can be found in \citet{Leconte:2013nat} (Supplementary Materials). 

Both CO$_2$ and H$_2$O cycles are included in the GCM used in this work. In our model, CO$_2$ can condense to form CO$_2$ ice clouds and surface frost if the temperature drops below the saturation temperature of CO$_2$ (at a given CO$_2$ partial pressure). A self-consistent water cycle is also included in the GCM. In the atmosphere, water vapour can condense into liquid water droplets or water ice particles, depending on the atmospheric temperature and pressure, forming clouds.

The fraction of cloud particles $\alpha_\text{c,liquid}$ (in $\%$) in liquid phase is given by \citep{Charnay:2014these}:
\begin{equation}
\alpha_\text{c,liquid}~=~\frac{T-(273.15-18)}{18}
\label{liq}
\end{equation}
where T is the atmospheric temperature of the corresponding GCM air cell. Above 0$^\circ$C, particles are fully liquid and below -18$^\circ$C they are assumed to be fully solid.

We used a fixed number of activated cloud condensation nuclei (CCNs) per unit mass of air $N_c$ to determine the local H$_2$O cloud particle sizes, based on the amount of condensed material. Following \citet{Leconte:2013nat}, we used $N_c$~=~10$^{4}$~kg$^{-1}$ for water ice clouds and 10$^{6}$~kg$^{-1}$ for liquid water clouds. These numbers - that give satisfactory results to reproduce the present-day Earth climate \citep{Leconte:2013nat} - are highly uncertain for post-impact conditions on Mars. 
On the one hand, impact events would inject a huge number of silicated particles in the atmosphere, potentially serving as CCNs. On the other hand, the huge rate of precipitation recorded in our 3-D simulations would remove efficiently these silicated particles. Eventually, we used $N_c$~=~10$^{5}$~kg$^{-1}$ for CO$_2$ ice clouds following \citet{Forget:2013}.

The effective radius $r_\text{eff}$ of the cloud particles is then given by:
\begin{equation}
r_\text{eff}~=~(\frac{3~q_c}{4 \pi \rho_c N_c})^{1/3}
\end{equation}
where $\rho_c$ is the density of the cloud particles (1000~kg~m$^{-3}$ for liquid 
and 920~kg~m$^{-3}$ for water ice) and $q_c$ is the mass mixing ratio of cloud particles 
(in kg per kg of air). The effective radius of the cloud particles is then used to compute both (1) 
their sedimentation velocity and (2) their radiative properties calculated by Mie scattering (see 
\citealt{Madeleine:2011these} for more details) 
for both liquid and ice cloud particles.

Water precipitation is divided into rainfall and snowfall, depending on the nature (and thus the temperature) 
of the cloud particles. Rainfall is parameterized using the scheme from \citet{Boucher:1995}, 
accounting for the conversion of cloud liquid droplets to raindrops by coalescence with other droplets. 
Rainfall is considered to be instantaneous (i.e. it goes directly to the surface) but can evaporate while 
falling through sub-saturated layers. The re-evaporation rate of precipitation $E_\text{precip}$ (in kg/m$^3$/s) is 
determined by \citep{Gregory:1995}:
\begin{equation}
E_\text{precip}~=~2 \times 10^{-5} (1-\frac{q_{v}}{q_{s,v}}) \sqrt{F_\text{precip}}
\end{equation}
where $q_{v}$ and $q_{s,v}$ are the water vapour mixing ratios in the air cell and at saturation, respectively. 
$F_\text{precip}$ is the precipitation flux (in kg/m$^2$/s). 
Re-evaporation of precipitation refers to rain that evaporates in the dry lower atmosphere before it reaches the ground.

Snowfall rate is calculated based on the sedimentation rate of cloud particles in the atmospheric layer. The sedimentation velocity of particles $V_\text{sedim}$ (in m/s) is assumed to be equal to the terminal velocity that we approximate by a Stokes law:
\begin{equation}
V_\text{sedim}~=~\frac{2 \rho_c g {r_\text{eff}}^2}{9 \eta}~(1+\beta K_n)
\end{equation}
where $\eta$ is the viscosity of atmospheric CO$_2$ (10$^5$~N~s~m$^{-2}$) and g the gravity of Mars (3.72~m~s$^{-2}$). (1+$\beta K_n$) is a 'slip-flow' correction factor \citep{Rossow:1978}, with $\beta$ a constant equal to $\frac{4}{3}$ and $K_n$ the Knudsen number that increases with decreasing atmospheric pressure.

While the internal and potential energy of condensates (water clouds, here) is accounted for in our convective moist 
adjustement scheme (see \citealt{Leconte:2013nat} lapse rate formulation), we acknowledge that we did not account for the potential, 
kinetic and internal energy carried by precipitation and that is dissipated in the atmosphere while falling and on the surface 
energy budget while reaching the surface. A possible strategy to implement the impact of precipitation on the 
surface energy budget is discussed in \citet{Ding:2016}.

At the surface, liquid water and water ice can co-exist. Their contributions are both taken into account in the albedo calculation as in \citet{Wordsworth:2013}.
The stability of liquid water/ice/CO$_2$ ice at the surface is governed by the balance between radiative, latent and sensible heat fluxes (direct solar insolation, thermal radiation from the surface and the atmosphere, turbulent fluxes) and thermal conduction in the soil. Melting, freezing, condensation, evaporation, and sublimation physical processes are all included in the model as in \citet{Wordsworth:2013} and \citet{Turbet:2017icarus}.


\subsection{1-D inverse climate model simulations}
\label{large_impact_1Dmodel_method}

The model described in this subsection was used to produce the results described in Section~\ref{impact_1D}.

Our 1D LMD Generic inverse\footnote{The inverse model does not solve for temperature, 
which is specified. Instead, it solves for the TOA fluxes in shortwave and longwave spectral ranges.} 
model is a single-column inverse radiative-convective climate model following the same approach ('inverse modeling') as \citet{Kasting:1984}, and using the same parameterizations as \citet{Ramirez:2017}. The atmosphere is decomposed into 200 logarithmically-spaced layers that extend from the ground to the top of the atmosphere arbitrarily fixed at 1~Pascal. The atmosphere is separated in three (at most) physical layers constructed as follows. First, we fix the surface temperature to the desired value. The first layer is constructed by integrating a moist (H$_2$O) adiabat upwards until CO$_2$ starts to condense. This first layer defines a convective troposphere assumed to be fully saturated. From the altitude where CO$_2$ starts to condense, we construct the second layer by integrating a moist (CO$_2$) adiabat upwards until the atmospheric temperature reaches the stratospheric temperature, arbitrarily fixed at 155~K as in \citet{Ramirez:2017}. 

Once the thermal profile of the atmosphere is constructed, we compute the radiative transfer 
(described in subsection~\ref{impact_rad_transfer}) in both visible and thermal infrared spectral domains, and through the 200~atmospheric layers. 
From this, we derive (1) the Outgoing Longwave Radiation (OLR) and (2) the planetary albedo, from which we can calculate the 
Absorbed Solar Radiation (ASR). Top Of Atmosphere (TOA) radiative budget can then be computed using OLR and ASR.
The radiative transfer calculations 
are described in details in subsection~\ref{impact_rad_transfer}.

Following \citet{Ramirez:2017}, we assumed that the planet is flat and the Sun remains fixed at a zenith angle of 60$^{\circ}$. The surface albedo is fixed to 0.216.

Our model can also take into account the radiative effect of clouds following the same approach as in \citet{Ramirez:2017}. A cloud layer can be placed at any arbitrary height (in any of the 200~atmospheric layers, and in any of the three physical layers previously described). 
We assume 1~km thick cloud decks as in \citet{Ramirez:2017}. Following \citet{Ramirez:2017}, we compute the optical depth $\tau_{\text{ice}}$ of the water ice clouds as follows:
\begin{equation}
    \tau_{\text{ice}}=\frac{3~Q_{\text{eff}}~IWC~\Delta z}{4~r_{\text{ice}}~\rho_{\text{ice}}},
\end{equation}
with $Q_{\text{eff}}$ the extinction coefficient, IWC the ice water content (in g/m$^3$), 
$\Delta z$ the vertical path length of the cloud layer, arbitrarily fixed to 10$^3$~m, 
$r_{\text{ice}}$ the effective radius of water ice particles, and $\rho_{\text{ice}}$ the volumetric mass of water ice, 
We used the same Mie optical properties (tabulated values of $Q_{\text{eff}}$) for the cloud particles 
as in the 3-D Global Climate Model (same radiative properties as used in \citealt{Wordsworth:2013}).
We assumed that the IWC scales following \citet{Ramirez:2017}:
\begin{equation}
    IWC~=~0.88~P,
\end{equation}
with P the atmospheric pressure at the cloud deck level. To explore the sensitivity of the results to the cloud content, we used the 'Relative Ice Water Content' which is a multiplicative factor applied to the IWC \citep{Ramirez:2017}. It is equal to 1 unless specified.

\subsection{Radiative transfer}
\label{impact_rad_transfer}

Our climate models include a generalized radiative transfer code adapted to any mixture of CO$_2$ and H$_2$O gases. Our radiative transfer calculations are performed on 38 spectral bands in the thermal infrared and 36 in the visible domain, using the 'correlated-k' approach \citep{Fu:1992} suited for fast calculations. 16 non-regularly spaced grid points were used for the g-space integration, where g is the cumulative distribution function of the absorption for each band.

Absorption caused by the absorption of H$_2$O and CO$_2$ in the atmosphere was computed using $kspectrum$~\citep{Eymet:2016} to yield high-resolution line-by-line spectra. We used the HITRAN2012 database for the H$_2$O and CO$_2$ line intensities and parameters \citep{Rothman:2013}. 
In addition, we incorporated the half-width at half maximum of H$_2$O lines broadened by CO$_2$ ($\gamma^{\text{H}_2\text{O}-\text{CO}_2}$) and CO$_2$ lines 
broadened by H$_2$O ($\gamma^{\text{CO}_2-\text{H}_2\text{O}}$), as well as the  corresponding temperature dependence exponents ( $n^{\text{H}_2\text{O}-\text{CO}_2}$ and $n^{\text{CO}_2-\text{H}_2\text{O}}$ ), 
based on \citet{Brown:2007,Sung:2009,Gamache:2016} and \citet{Delahaye:2016}. More details can be found in \citet{Turbet:2017LPI} and \citet{Tran:2018}.

Collision-induced absorptions, dimer absorptions and far wing absorptions were also taken into account, 
whenever data was available. Far wings of CO$_2$  band lines 
(both CO$_2$-CO$_2$ and CO$_2$-H$_2$O) were computed using the $\chi$-factor approach, 
using experimental data from \citet{Perrin:1989}, \citet{Tran:2011} and \citet{Tran:2018}. 
The $\chi$-factor is an empirical correction of the Lorentzian line shape adjusted to laboratory measurements. 
CO$_2$-CO$_2$ collision-induced and dimer absorptions were computed based on \citet{Gruszka:1997,Baranov:2004,Stefani:2013}. 

H$_2$O-H$_2$O continuum was taken into account using the MT$\_$CKD~3.0 database \citep{Mlawer:2012}, 
from 0 to 20,000~cm$^{-1}$. MT$\_$CKD databases are available on \url{http://rtweb.aer.com/}. H$_2$O-CO$_2$ 
continuum was calculated with the line shape correction functions digitized from \citet{Ma:1992} 
(validated experimentally by \citealt{Tran:2019}) using line positions and 
intensities from the HITRAN2012 database \citep{Rothman:2013}, with a cut-off distance at 25~cm$^{-1}$, and from 0 to 
20,000~cm$^{-1}$.  The temperature dependence of the continuum was empirically derived using data 
digitized from \citet{Pollack:1993}. 

More details on the CIAs, dimer absorptions and far wing absorptions can be found in \citet{Turbet:2017LPI} and \citet{Tran:2018}.

\subsection{2-D Mantle Dynamics model simulations}
\label{large_impact_2D_mantle_dynamics_model_method}

The model described in this subsection was used to produce the results described in Section~\ref{stagyy_part}.

We used the StagYY 2-D mantle dynamics code \citep{Tackley:2008} to simulate the long-term effects of very large impacts on 
the interior of Mars. Specifically, we used the version of the
code from \citet{Armann:2012} and \citet{Gillmann:2014},
adapted to Mars conditions. Geometry is set to 2-D spherical annulus
with a 512~x~64 grid and 2 million advected tracers. 

The equations of mass, energy and momentum conservation are solved in the mantle, that is considered anelastic, 
compressible and to be described by the infinite Prandtl number approximation. 
Boundary conditions at the top of the mantle are free-slip. Top temperature is set by surface temperature. 
Bottom temperature decreases with core temperature following \citet{Nakagawa:2004}. Physical properties like
density, thermal expansivity, and thermal conductivity are depth
dependent and are calculated as described in \citet{Tackley:1996}. 
The rheology is temperature- and pressure-dependent diffusion creep. 
Activation energy and volume are chosen according to \citet{Karato:2003} for dry olivine. 
Initially, radiogenic elements are
uniformly distributed. Radiogenic heating decreases with time and is
treated as in \citet{Armann:2012}. The mineral solid-phase
transitions in the olivine system and in the pyroxene-garnet system are
included as discussed in \citet{Xie:2004}. Other physical,
Mars-specific parameters used in the convection simulations are
detailed in \citet{Keller:2009}. 
Here we model only solid state convection of the mantle, in a similar way to what has been used previously in studies 
of impact heating on Mars \citep{Roberts:2009,Roberts:2014}.
Melting is treated as in other
studies using the StagYY code \citep{Nakagawa:2004,Xie:2004,Armann:2012}. 
Only melt located above a
certain depth (set to~600 km) is considered to be positively buoyant
\citep{Ohtani:1998}. We assumed it to instantaneously erupt at the
surface \citep{Reese:2007}, since its migration can be considered to
be fast compared to convection processes. Extracted melt is emplaced at
the surface at surface temperature. On the timescales relevant for
mantle dynamics evolution, surface conditions are assumed not to vary;
simulations have been run for constant surface temperatures from 200K to
300K, without significant modification of the convection pattern. Full
coupling featuring in \citet{Gillmann:2014} and \citet{Gillmann:2016} is not used here. Atmosphere content is tracked to assess
volatile degassing and constrain simulations based on present-day
observation.
Impacts are treated in the same way as in \citet{Gillmann:2016}. 
Impacts are assumed to be vertical (head-on)
for the sake of mantle dynamics effects modelling. 
Velocity of the impactor was set to 9 km$/$s for the reference impact. Reference impact
time is 4~Ga before present; alternative simulations with different times
have been run (4.2 to 3.8 Ga) without significant change. The projectile
is assumed to be an asteroid of density 3000~kg$/$m$^3$. Volatile composition
can vary widely depending on the type of bolide impactors. 

Three effects of impacts are considered: atmospheric erosion, volatile
delivery and heat deposition in the mantle. Atmospheric erosion has been
calculated from parameterization based on the SOVA hydrocode simulations
\citep{Shuvalov:2009,Shuvalov:2014} and extrapolated for larger
bodies when necessary, as described by \citet{Gillmann:2016}. An upper
limit to erosion is set by the tangent plane model \citep{Vickery:1990}.
Erosion of the atmosphere is found to be limited to 0.1-1$\%$ of the total
atmosphere at the time of impact for the range of impactors considered.
It has thus a minor effect on the global evolution, especially as it is
countered by volatile deposition. This second mechanism primarily
depends on projectile volatile content. It is limited by impactor mass
loss as ejecta during impact process \citep{Shuvalov:2009} and by the
portion of the projectile avoiding melting \citep{Svetsov:2015}.
For large bolide impactors, most of the volatile constituents ($\sim$~80$\%$) are
delivered to the atmosphere. 
Effects of the impact on the mantle are treated as in \citet{Gillmann:2016}. 
Shock pressure, generated by high energy collisions, followed by
adiabatic decompression generates heat, leading to the formation of the
so-called thermal anomaly in the mantle. A parameterized law from
\citet{Monteux:2007} is used to account for heat emplaced in the
martian mantle on collision. Accretion of solid material and crater
excavation are not considered. We neglect crust vaporization by the impactor. Thermal anomalies are integrated into the StagYY
temperature field, produce melting and affect subsequent convection
patterns.

\section{Exploration of the environmental effects of a very large impactor with a full 3-D Global Climate Model}
\label{impact_3D}

Here we study the environmental effects of very large impact events on the atmosphere and surface of early Mars, 
using the 3-D Global Climate Model presented in Section~\ref{large_impact_3Dmodel_method}. 
Because Global Climate Model simulations are computationally very expensive, we focus in this section on one reference 
post-impact simulation (computational cost of $\sim$~50,000~CPU hours on the 
French OCCIGEN supercomputer) of a very large impactor hitting the surface of Mars, initially assumed to be 
endowed with a 1~bar pure CO$_2$ atmosphere. 
This value of CO$_2$ surface atmospheric pressure was arbitrarily chosen, but is roughly consistent with the various 
mineralogical and isotopic constraints summarized in \citealt{Kite:2019} (Figure~9). 
We explore the effect of surface pressure with a computationnally much more efficient 1-D numerical climate model in Section~\ref{impact_1D}.
The impactor is assumed to be large enough to trigger the 
vaporization of $\sim$~2~bar (i.e. 54~m GEL) of water into the atmosphere
\footnote{We explore in Sections~\ref{impact_3D} and \ref{impact_1D} how the initial water and CO$_2$ content affect the 
impact-induced climate change.}. The atmosphere, surface, and 
subsurface (down to $\sim$~80~m, i.e. the deepest layers in the model) are 
assumed to be suddenly and uniformly heated up to 500~K\footnote{Although the choice of the post-impact 
temperature seems arbitrary here, we show in Section~\ref{large_impact_chronology} 
that the impact-induced climate state becomes independent of initial post-impact temperature 
from the moment the first droplet of liquid water hits the ground. 
The temperature is then controlled by the condensation temperature of the water vapour 
(determined by the Clausius-Clapeyron equation) and not the initial post-impact temperature.}. 

Such extreme post-impact conditions 
are likely typical of the few most extreme impact events forming the largest basins observed on Mars \citep{Segura:2002,Toon:2010}.

\subsection{Chronology of the event}
\label{large_impact_chronology}

\begin{figure*}
\centering
\includegraphics[scale=0.15]{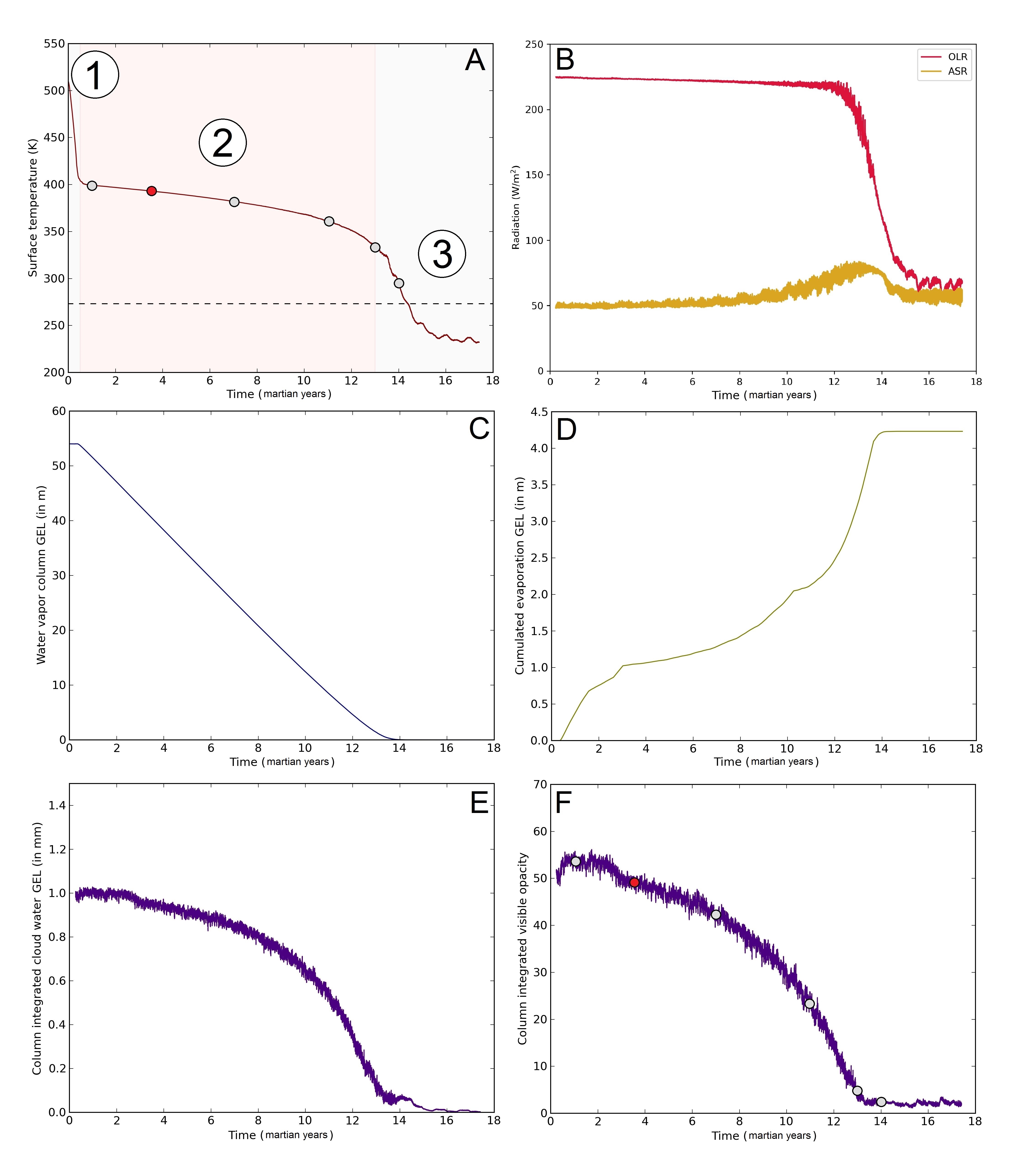}
\caption{Globally averaged temporal evolution of the (A) surface temperature (in K), (B) top of atmosphere radiative fluxes 
(in W/m$^2$), (C) integrated column of water vapour (in m GEL), (D) cumulative surface evaporation of water (in m GEL), 
(E) column integrated cloud water content (in mm GEL) and (F) column integrated cloud visible opacity. All these quantities 
were computed for the reference simulation described extensively in Section~\ref{impact_3D} of this paper. The grey/pink zones 
(and numbers {1,2,3}) in panel A indicate the three first post-impact phases described in Section~\ref{large_impact_chronology}.}
\label{large_impacts_1D_mean_plots}%
\end{figure*}

Our 3-D simulations indicate that the sequence of events following the very large impact event previously described can be decomposed into four main phases:

Phase~I: very hot atmosphere, no precipitation. In this first phase, the atmosphere is too warm for water 
to condense on the surface. Precipitation (produced in the cooler, upper atmosphere) is 
re-evaporated (in the lower atmospheric layers). This phase ends when the first droplet of water 
reaches the ground. The atmosphere is then almost entirely saturated in water vapour, and the atmospheric 
thermal profile follows a moist adiabat as shown in Fig.~\ref{large_impacts_profiles_1D-3D}A. The 
atmospheric state (i.e. mean thermal structure) at the end of this first phase (i.e. when the first droplet of water reaches the ground) 
depends on the amount of CO$_2$ and H$_2$O in the atmosphere, but does not depend on the initial 
post-impact temperature assumed. This means that, for the impact event described here (1~bar CO$_2$ 
atmosphere + 2~bar of water vaporized), the initial temperature (here, 500~K) does not have any major 
effect on the nature of the environmental effects (on the atmosphere and the surface, but not the subsurface) 
of the impact during the following phases (phases II, III and IV described below). 
The duration of this first phase is roughly controlled by (i) the net Top Of Atmosphere (TOA) radiative 
budget and (ii) the amount of extra thermal energy of the atmosphere (i.e. the difference of thermal energy 
between the initial post-impact thermal profile - here a 500~K isotherm thermal profile - and a moist adiabatic thermal profile.). 
The duration of this phase 
is usually short because the initially hot atmosphere quickly cools by emitting thermal radiation to space. 
For our reference simulation, it takes $\sim$~0.5~martian year\footnote{One martian year lasts 
approximately 687~Earth days, i.e. 1.88 Earth year.} for the first droplet of water to reach the surface, 
which sets the end of this first phase (see Fig.~\ref{large_impacts_1D_mean_plots}A). This phase 
is not interesting from the point of view of surface erosion because no water is present at the surface. 


Phase~II: hot atmosphere, intense precipitation. This second phase starts when the atmosphere becomes 
almost entirely saturated in 
water vapour, and water can start to rain on the surface and accumulate. This phase ends when most water vapour has 
condensed on the surface. This is the most interesting phase because it coincides with the main bulk of precipitation (rainfall). During 
this phase, our 3-D Global Climate simulations indicate that (1) a thick, reflective and quasi-uniform cloud cover 
is present and (2) the net radiative budget at the Top Of the Atmosphere (TOA) is roughly constant, because both the 
outgoing longwave radiation (OLR) and the planetary albedo are constant (see Fig.~\ref{large_impacts_1D_mean_plots}B). 
For our reference simulation, the net TOA radiative budget is -180~W/m$^2$ (indicating the planetary atmosphere 
must cool to reach equilibrium). 
As a result, the water 
vapour atmospheric content gets progressively depleted (see Fig.~\ref{large_impacts_1D_mean_plots}C). The duration of 
this second phase is roughly controlled by (i) the net TOA radiative budget and (ii) the total amount of latent heat 
that can be extracted from the condensation of the entire water vapour atmospheric column. For our reference simulation, 
the duration of the phase is $\sim$~12~martian years. This duration can be empirically approximated by 
$q_{\text{col,H}_2\text{O}} (m)/4.5$, where $q_{\text{col,H}_2\text{O}}$ is the initial global mean integrated 
column of water vapour in GEL (m). Approximately 2.6~m of water condenses on the surface per Earth year, which is 
very similar to the result obtained by \citet{Segura:2002} with a 1-D cloud-free numerical climate model. This rainfall rate is 
constant during phase~II (because the net TOA desequilibrium is constant) indicating that the rainfall rate does not depend 
on the initial water inventory (as long as water is a radiatively dominant atmospheric species).
Note that the surface evaporation of water is very limited during this phase. In our reference simulation, approximately 8$\%$ 
of the precipitation gets re-evaporated from the surface (see Fig.~\ref{large_impacts_1D_mean_plots}D). We investigate 
this phase in more details in the next subsections.

Phase~III: conversion of surface liquid water into ice. When the third phase starts, most water vapour 
has already condensed on the surface. Surface water rapidly freezes and the planet gets cold 
(see Fig.~\ref{large_impacts_1D_mean_plots}A), even colder than before the impact event because the 
planet is now covered by a thick, reflective ice cover (this ice cover is actually expected to be thicker in the topographic lows,
i.e. basins and the northern lowlands where liquid water would flow during phase~II). Based on the results of \citet{Turbet:2017icarus}, 
water - that should accumulate in the topographic depressions of the planet - 
would freeze (down to the bottom) in 10$^3$-10$^5$ years maximum (depending on the total water content).

Phase~IV: migration of water ice to the cold traps. In the fourth phase, water progressively migrates to the cold traps of the planet. 
For our reference simulation (high CO$_2$ atmospheric pressure, high obliquity), 
water should migrate to the southern highlands \citep{Wordsworth:2013,Bouley:2016,Turbet:2017icarus}. 
We do not explicitly simulate this phase here. However, based on the results of \citet{Turbet:2017icarus}, 
water should migrate to the cold traps of the planet within 10$^4$-10$^6$ years (depending on the total water content).

\medskip

In total, it takes $\sim$15~martian years for the surface temperature to drop below the freezing temperature of water (273~K). 
Note that subsurface temperatures could remain above 273~K for much longer periods of time, 
as reported in \citet{Segura:2002}. At the end of our simulations (after $\sim$~18~martian years), 
the mean (regolith) subsurface temperatures at 5 and 50~m are respectively 320 and 470~K.

\medskip

\cite{Steakley:2019} recently independently presented a similar chronology of events (also in four distinct phases) derived 
from 3-D Global Climate Model simulations of smaller impactors.

\medskip

Below we investigate in much more details the second phase, because this is the phase in which most of the precipitation (rainfall) occurs.

\medskip

\begin{figure}[htbp] 
 \centering
\includegraphics[scale=0.15]{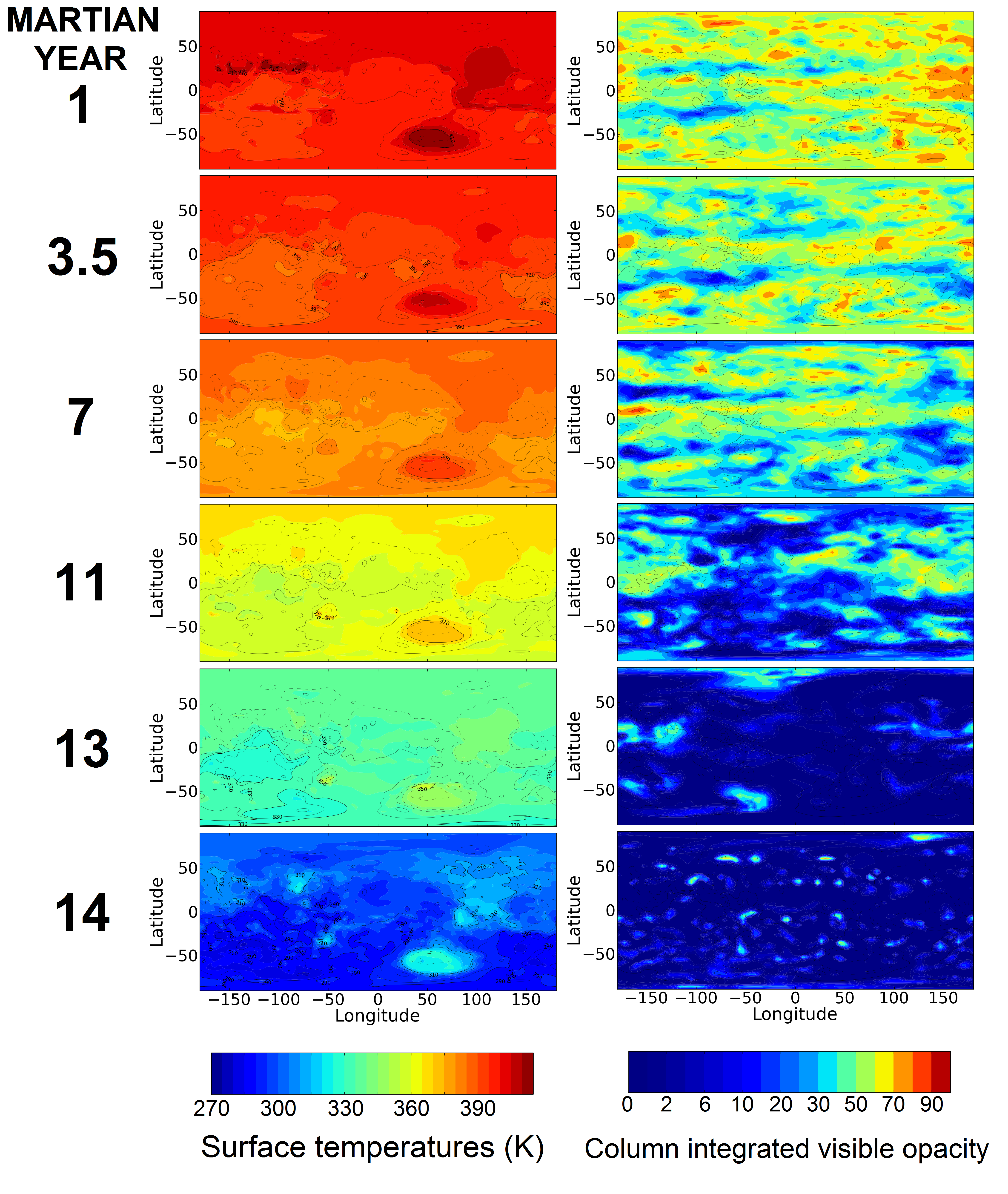}
\caption{Snapshots of post-impact surface temperature (in K) and visible opacity maps at six distinct timings (1, 3.5, 7, 11, 13 and 14 martian years). These six timings are also highlighted in Fig.~\ref{large_impacts_1D_mean_plots}AF with dots.}
 \label{large_impacts_evolution_maps}
\end{figure}

\subsection{Clouds and radiative budget}
\label{large_impact_clouds}

\begin{figure}[htbp] 
 \centering
\includegraphics[scale=0.22]{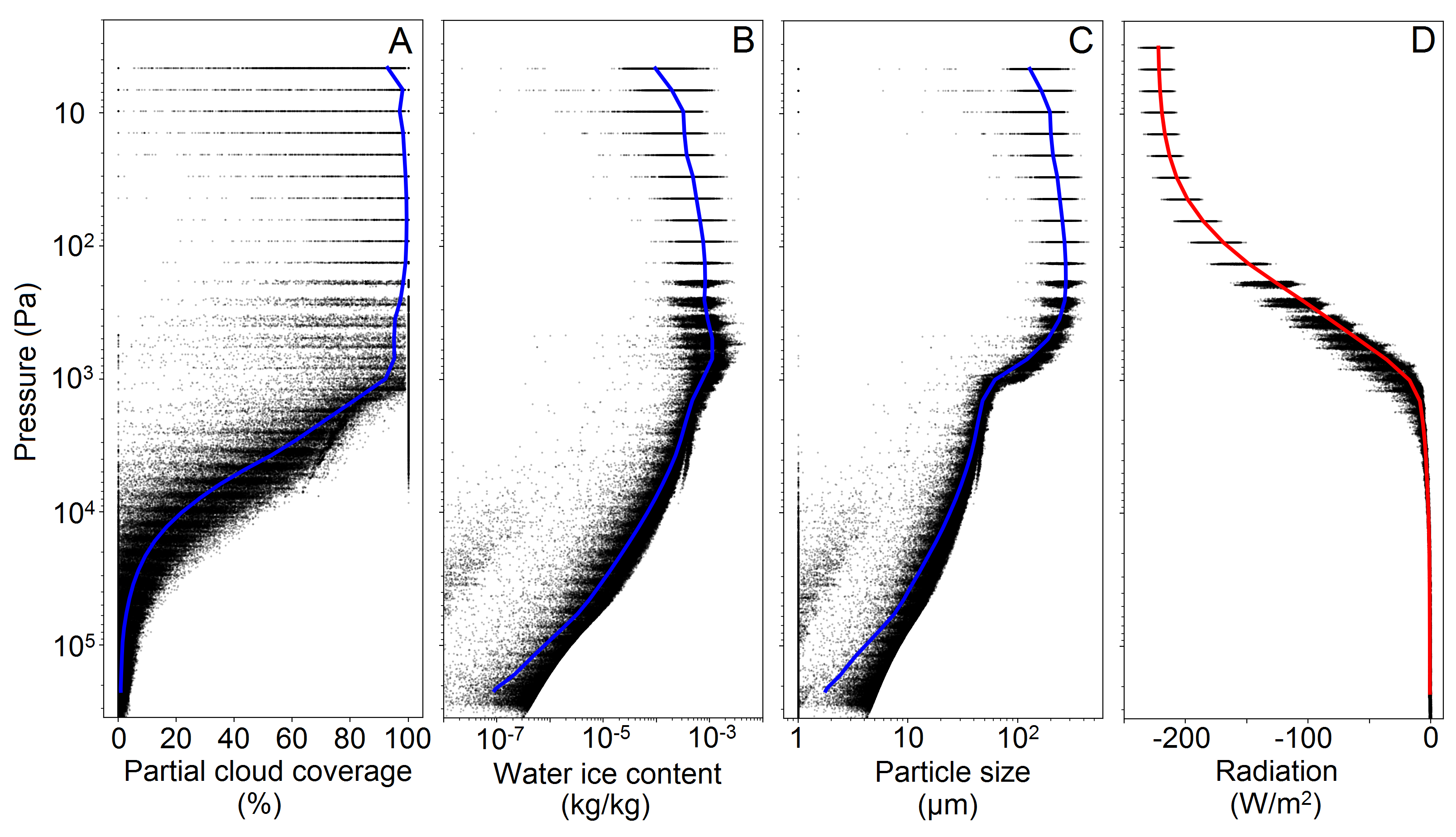}
\caption{Vertical profiles of (A) the partial cloud coverage (in $\%$), (B) the water ice content (in kg/kg), 
(C) the effective radius of cloud particles (in $\mu m$) and (D) the net thermal 
infrared flux passing through each of the atmospheric layer (negative means upward). 
These profiles are snapshots calculated 3.5~martian years after the impact event. 
This timing is highlighted in Fig.~\ref{large_impacts_1D_mean_plots} with a red dot. 
The colored thick lines vertical profiles are globally averaged snapshots. Small black 
dots are snapshots of all possible values reached by GCM air cells. In total, there are 
276480 (i.e. the number of GCM air cells) black dots in each subfigure. 
Note that comb-like structures appear because 
fixed pressure levels were used in the upper atmosphere (see Section~\ref{large_impact_3Dmodel_method}).}
 \label{large_impacts_profiles_cloud_emission}
\end{figure}

The second phase starts when the first droplet of (liquid) water reaches the surface of Mars. At this stage, the atmosphere is almost entirely saturated in water vapour. The outgoing longwave radiation (OLR) is roughly constant through time (see Fig.~\ref{large_impacts_1D_mean_plots}B) and across the planet. The OLR is dominated by the thermal infrared emission of the moist, upper atmosphere (see Fig.~\ref{large_impacts_profiles_cloud_emission}D). This result is similar in nature with the asymptotic behaviour of the OLR predicted by 1-D radiative convective models assuming a thermal profile following the water vapour saturation curve \citep{Nakajima:1992,Kasting:1993,Kopparapu:2013}. This is the typical state reached by planets entering in runaway greenhouse. For a Mars-size planet, \citet{Kopparapu:2014} estimates that the asymptotic OLR (for a pure water vapour atmosphere) is 250~W/m$^2$, i.e. 20~W/m$^2$ higher than our result. This small difference is likely due to (i) a different CO$_2$ atmospheric content (discussed in the next section ; see also \citealt{Goldblatt:2012,Ramirez:2014b} and \citealt{Marcq:2017}), (ii) different treatments of water vapour absorption \citep{Kopparapu:2013}, (iii) the radiative effect of clouds in the infrared.

\medskip

Thermal radiation cooling occurs mainly in the upper atmospheric layers 
(see Fig.~\ref{large_impacts_profiles_cloud_emission}D). This triggers moist convection 
and thus water vapour condensation, forming clouds in the upper atmosphere (see Fig.~\ref{large_impacts_profiles_cloud_emission}ABC). 
Because this radiative cooling occurs everywhere (at each latitude and longitude of the planet), 
clouds form everywhere on the planet (see Fig.~\ref{large_impacts_evolution_maps} and \ref{large_impacts_profiles_cloud_emission}A). 
This result is qualitatively in agreement with \citet{Segura:2008}. Moreover, simple energy conservation consideration 
(supported by our 3-D Global Climate simulations) show that approximately 1kg/m$^2$ of cloud particles are produced every 3 hours. 
The production rate of cloud particles is so high that cloud particles accumulate, growing to large sizes up to several hundreds of 
microns for icy particles in the upper cloud layer (see Fig.~\ref{large_impacts_profiles_cloud_emission}C). The accumulation of cloud 
particles is limited by (i) coagulation of cloud liquid droplets into raindrops (following the numerical scheme of \citealt{Boucher:1995}) 
and (ii) sedimentation of large ice particles (parameterized following a Stokes law \citep{Rossow:1978}). As a result, our 3-D Global 
Climate simulations show that a thick, uniform cloud cover is produced in the upper layers of the planet 
(see Fig.~\ref{large_impacts_profiles_cloud_emission}ABC).

This thick, uniform cloud cover (mostly located in the upper atmosphere, as illustrated in Fig.~\ref{large_impacts_profiles_cloud_emission}AB) reflects incoming solar radiation efficiently. In average, the planetary albedo reaches $\sim$~0.55 (see Fig.~\ref{large_impacts_1D_mean_plots}B). Moreover, a large fraction of the incoming solar radiation is absorbed in the upper atmospheric layers, mostly by water vapour and clouds. As a result, the strong deficit of absorbed solar radiation versus outgoing longwave radiation (see Fig 1D) cools down the planet very rapidly. During most of this phase, the radiative disequilibrium at the top of the atmosphere is $\sim$~-180W/m$^2$. As a result, the atmosphere and the planet progressively cool down (see Fig.~\ref{large_impacts_1D_mean_plots}A and \ref{large_impacts_evolution_maps}).

\medskip

As the planet cools down, the globally averaged cloud water content and visible 
opacity slowly decrease (see Fig.~\ref{large_impacts_1D_mean_plots}EF). While water is progressively 
raining out from the atmosphere, the atmosphere is progressively drying out. The upper atmosphere become slowly 
drier through time, producing more and more unsaturated regions. Progressively, we enter in a regime of radiative fins 
as predicted by \citet{Pierrehumbert:1995} and simulated with a 3-D Global Climate Model in \citet{Leconte:2013nat} in the 
context of the runaway greenhouse, where the emission can locally exceed
the maximum emission for a saturated atmosphere. 
This can be observed in Fig.~\ref{large_impacts_evolution_maps} (right panel) where the 
clouds become more and more patchy through time. Regions where clouds are absent coincide (i) with sub-saturated regions and (ii) 
with regions where the thermal emission to space exceeds the maximum emission ($\sim$~230W/m$^2$, here) calculated for a quasi-saturated 
atmosphere.

\medskip


\begin{figure*}
\centering
\includegraphics[scale=0.25]{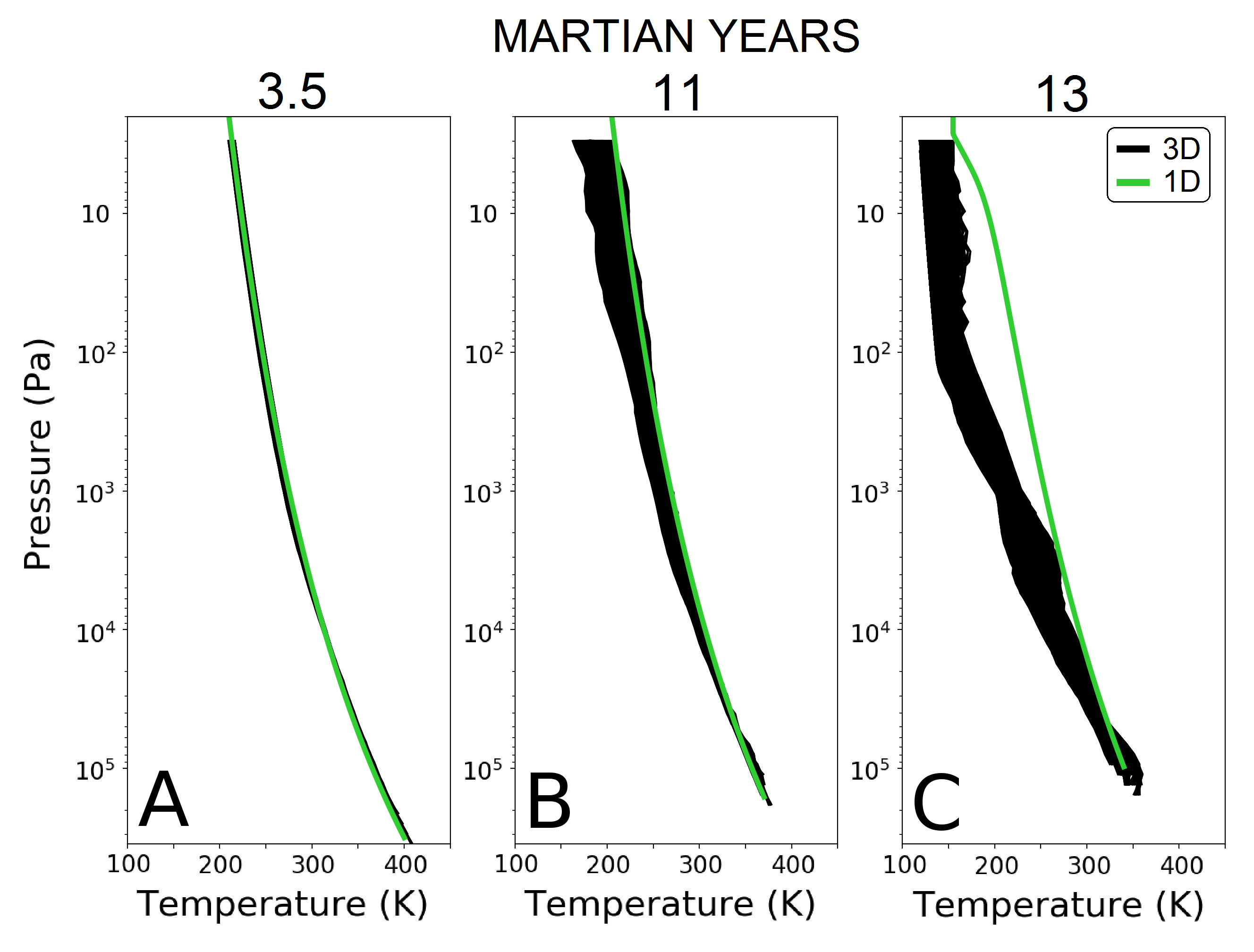}
\caption{Snapshots of the vertical thermal profiles after 3.5 (left panel), 11 (middle panel) and 
13~martian years (right panel). The black region indicates all the $\{$pressure,temperature$\}$ reached 
in the 3-D Global Climate simulation. The green solid line indicates the thermal profile calculated in our 
1-D inverse climate model (see description in Section~\ref{large_impact_1Dmodel_method}) for the 
same surface temperature and CO$_2$ partial pressure as assumed in the GCM reference simulation.}
\label{large_impacts_profiles_1D-3D}%
\end{figure*}

While the planet cools down, not only the globally averaged surface temperature decreases, 
but also the variability of the surface temperature and more generally of the atmospheric temperature 
progressively increases (see Fig.\ref{large_impacts_profiles_1D-3D}ABC). During most of the second phase, 
the 1D thermal (and water vapour, respectively) profiles calculated at each location of the planet 
in the GCM follow remarkably well the thermal (and water vapour, respectively) 
profile predicted by 1D climate calculations (see Fig.~\ref{large_impacts_profiles_1D-3D}A) assuming a fully saturated 
profile. But as the planet cools down, subsaturated regions appear and discrepancies with the 1-D calculations start to emerge 
(see Fig.~\ref{large_impacts_profiles_1D-3D}BC). The fact that the thermal and water vapour profiles match very well 1-D calculations during 
most of the second phase is important, because it indicates that we can use a 1-D model 
(computationally much more efficient than 3-D simulations) to 
explore the nature of the post-impact main phase of precipitation, depending on many different parameters (e.g. CO$_2$ and H$_2$O 
total atmospheric contents). This exploration is the topic of Section~\ref{impact_1D}.

\subsection{Localization of precipitation}

\begin{figure*}
\centering
\includegraphics[scale=0.32]{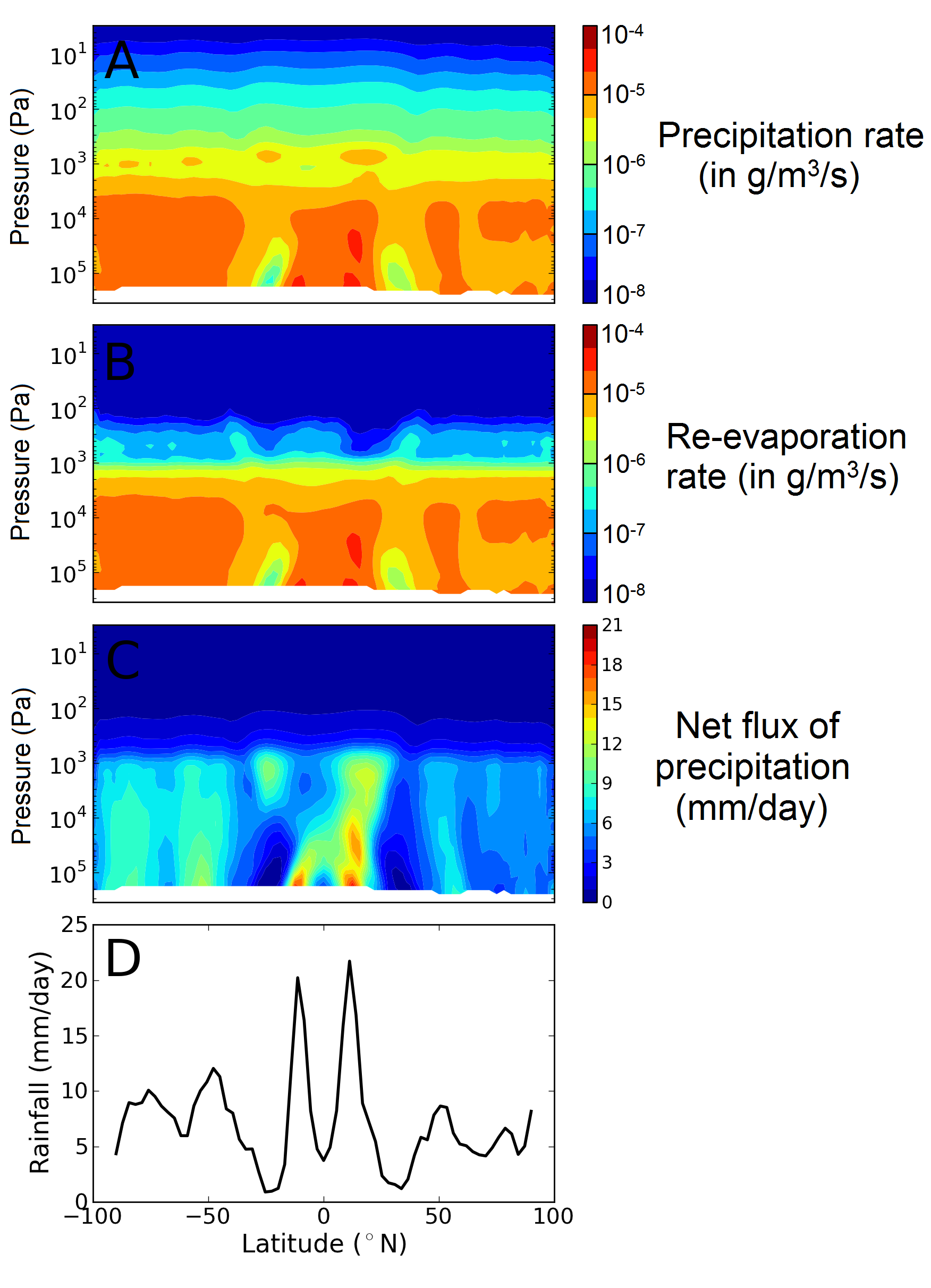}
\caption{Snapshot of the zonal means of (A) the cloud particle rate of precipitation (in g/m$^3$/s), (B) the rate of re-evaporation of precipitation (in g/m$^3$/s), (C) the downward flux of precipitation (accounting for re-evaporation) in kg/m$^2$/day (or mm/day) and (D) the surface accumulation of precipitation (in mm/day). This is a snapshot 3.5~martian years after the reference impact event.}
\label{large_impacts_budget_precipitation}%
\end{figure*}

Precipitation is produced uniformly in the upper cloud layer. Yet, most of the precipitation of the main, 
upper cloud layer is re-evaporated while falling through some sub-saturated lower layers (following the numerical scheme 
of \citealt{Gregory:1995}). This is illustrated in Figure~\ref{large_impacts_budget_precipitation} that presents a snapshot of 
the zonal mean budget of precipitation/re-evaporation. In fact, 3-D GCM simulations indicate that almost none of the 
precipitation produced in the upper cloud layer actually reaches the surface (see Fig.~\ref{large_impacts_budget_precipitation}). 
Instead, this is the condensation produced by the large scale air movements in the lower atmosphere that is the primary source of 
precipitation reaching the ground of the planet.

The equatorial regions receive (in our 3-D simulations) in average a few tens of W/m$^2$ of solar radiation in excess compared to 
the poles. This is likely enough to trigger large scale movements in the lower atmosphere, in particular to transport energy 
from the equator to higher latitudes, following a Hadley cell-like structure. Near the equator, ascending air produces condensation 
and thus precipitation (see Fig.~\ref{large_impacts_budget_precipitation}D). Near 30$^\circ$S/30$^\circ$N latitudes, this is the 
descending branch of the Hadley cell. These regions of air subsidence are noticeably subsaturated, and (almost) no precipitation 
reaches the ground (see Fig.~\ref{large_impacts_budget_precipitation}D). 

\begin{figure*}
\centering
\includegraphics[scale=0.24]{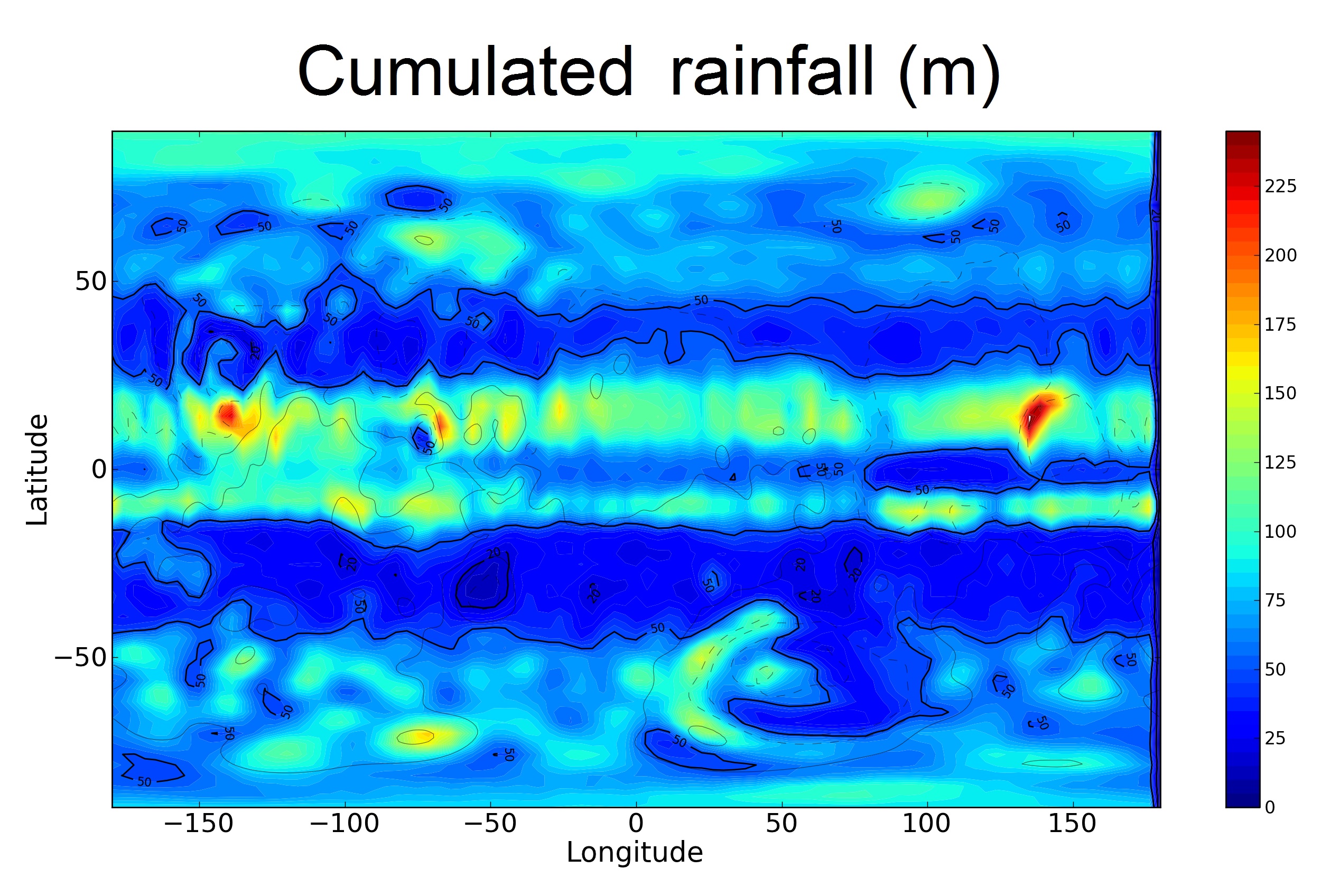}
\caption{Cumulative rainfall map (in m) 15 martian years after the bolide impact event. Ascending branches of 
the Hadley cells lie near the equator, 
while descending branches are located around $\pm$~30$^\circ$~latitude. 
The absence of rain at the equator is due to the presence of equatorially trapped modes 
producing westerly winds.}
\label{large_impacts_rain_map}%
\end{figure*}

Figure~\ref{large_impacts_rain_map} illustrates the fact that the precipitation (rainfall) patterns are mostly produced in response to 
the large scale circulation in the lower atmosphere. We note a peak of precipitation near the equator (ascending branch of Hadley cell) and 
a lack of precipitation near 30$^\circ$S/30$^\circ$N (descending branch of Hadley cell). 
As a result, the precipitation pattern recorded in the 3-D simulation is intriguingly anti-correlated 
with the position of the valley networks (located around 30$^\circ$S)
mapped in the pre-True Polar Wander topography \citep{Bouley:2016}. Future work could better explore how 
the latitudinal distribution of incoming solar radiation (e.g. due to different obliquity) or other source of forcing can 
impact the lower atmospheric circulation and thus the precipitation patterns.
Significant precipitation is also 
recorded at higher latitudes, likely produced by large scale atmospheric circulation. In addition, we also note some localized 
sources of precipitation (e.g. east of Hellas crater) or absences of precipitation (e.g. bottom of Argyre crater) 
likely due to coupling between lower atmospheric circulation and topography (e.g. through adiabatic cooling on the crater slopes, 
or intense re-evaporation of precipitation at the bottom of deep craters). 
Our 3-D numerical simulations suggest though that the topographic effect of large basins (e.g. Hellas, Argyre, Isidis) 
has a minimal contribution to the large-scale-circulation-induced precipitation patterns discussed here.

In average, $\sim$~58~m of precipitation (54~m from the initial water vapour reservoir, 4~m from 
evaporation of precipitation that reached the ground) 
is recorded throughout the 3-D simulations. Note that, although there are some latitudinal 
differences in the surface distribution of precipitation, cumulative precipitation is high everywhere 
on the planet (at least 20~m of cumulative precipitation). Precipitation is deluge-like with an average of 
2.6~m of precipitation per Earth year. 
Such heavy rains are incompatible with the formation of martian valley networks \citep{Barnhart:2009}. 
However, they could produce overland flow, which could be responsible for Noachian crater degradation \citep{Palumbo:2018impact}.


Moreover, because most of the precipitation that reaches the ground 
does not get re-evaporated, the impact-induced hydrological cycle is a 'one-shot' cycle 
(i.e. there is a low amount of hydrological cycling of water). 
The low amount of surface evaporation is due to the fact that the lower atmosphere 
is close to saturation during the second phase (hot atmosphere, intense precipitation) of the impact event. 
We acknowledge that we did not take into account the possibility that water could be transported from wetter 
to drier latitudes possibly through river flow, which may increase surface evaporation in the driest regions 
of the planet (e.g. near $\pm$~30$^{\circ}$ latitude). However, we believe this should not significantly increase 
the total surface re-evaporation of precipitation, because the lower atmosphere is 
close to saturation even in the driest regions of the planet.
At first order, the total amount of 
rainfall produced in response to a very large impact event can be approximated by the total amount of water vapour 
initially vaporized and/or sublimed in the atmosphere, i.e. by the total (near-surface,surface) Noachian water content.
Thus, the amount of rainfall produced by basin-scale impact events is orders of magnitude lower 
than that required for valley network formation \citep{Luo:2017,Rosenberg:2019}, 
even when accounting for all observed Noachian-aged basins.




\section{Exploration of the diversity of post-impact atmospheres with a 1-D inverse climate model}
\label{impact_1D}

3-D simulations are great to explore in details the post-impact atmospheric evolution due to atmospheric circulation, 
formation and evolution of clouds, etc. but are not suited (due to their high computational cost) to explore the sensitivity 
of the results to a wide range of parameters. We demonstrated in the previous section that the thermal profiles of the atmosphere 
in the GCM are well reproduced by 1D radiative-convective calculations during the main post-impact phase of precipitation 
(see Fig~\ref{large_impacts_profiles_1D-3D}AB). Here we use a 1-D radiative-convective inverse climate model presented in 
Section~\ref{large_impact_1Dmodel_method} to explore the sensitivity of the results obtained with our 3-D Global Climate Model. 

The initial post-impact CO$_2$ and H$_2$O atmospheric reservoirs are highly uncertain because 
they depend on the pre-impact CO$_2$ atmospheric pressure 
and (near-surface,surface) water reservoirs, as well as the amount 
and nature of volatiles delivered and excavated by the impactor\footnote{It has been recently proposed \citep{Haberle:2017h2} that 
fast thermochemical reactions that take place in the very hot post-impact martian atmosphere could produce H$_2$ and CH$_4$ 
that could generate afterwards an efficient CIA-induced greenhouse effect \citep{Wordsworth:2017,Turbet:2019icarus}. 
This effect is not modelled here, but 
deserves further investigations.}.
The water cloud properties (size of particles, thickness of clouds) are also highly uncertain because 
they depend on many exotic physical processes that are not properly modelled in the GCM. 
Thus, we want to explore how (i) the total CO$_2$ and H$_2$O initial inventories and (ii) the microphysics of clouds 
can affect the TOA (Top of Atmosphere) radiative budget and thus the duration of the impact-induced climate change. 


\subsection{Results from cloud-free numerical climate simulations}

\begin{figure*}
\centering
\includegraphics[scale=0.17]{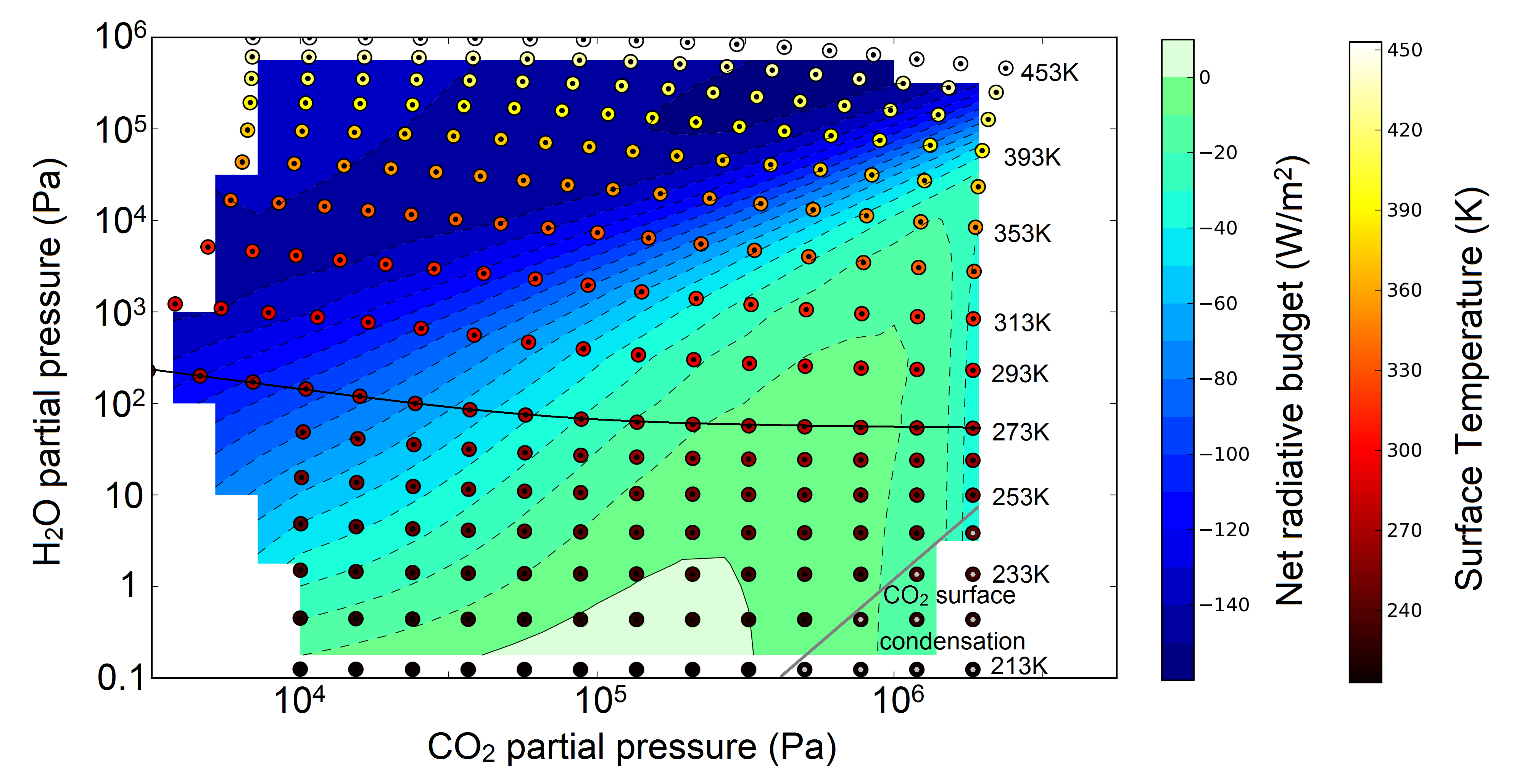}
\caption{Contour plot of the net radiative budget at the TOA (Top Of Atmosphere) of the post-impact early Mars atmosphere, 
as a function of CO$_2$ and H$_2$O partial pressures in the atmosphere. The contour plot was calculated 
by interpolating the data points of the scatter plot. The color of the data points corresponds to the surface 
temperature of the simulation. The planet is at equilibrium if the radiative budget is equal to 0, 
which is never the case for any of the post-impact simulations. 
For a given CO$_2$ atmospheric content, the increase in H$_2$O content leads to a decrease in the net TOA radiative budget 
despite the greenhouse effect of water vapour. This is due to the fact that the surface temperature also increases, which increases the OLR 
and decreases the net radiative budget at the TOA.
This dependency stops when the OLR reaches its asymptotic value.}
\label{large_impacts_scatter_plot}%
\end{figure*}

We first explore cloud-free numerical climate simulations to investigate the role of the initial, 
post-impact CO$_2$ and H$_2$O atmospheric inventories. Here we vary two parameters: the initial surface 
temperature (from 213 to 453~K) and the initial atmospheric pressure (from 3~mbar to 30~bar). Cold temperatures 
($<$~240~K) are typical of global mean pre-impact surface temperatures calculated for CO$_2$-dominated atmospheres 
\citep{Forget:2013,Wordsworth:2013}. Warm temperatures ($>$~240~K) are typical of post-impact surface temperatures expected 
during the main phase of precipitation (corresponding to the second phase depicted in Section~\ref{large_impact_chronology}). 
As a comparison, we recall that the maximum post-impact temperature recorded in our 3-D reference simulation (discussed 
in the previous section) during the precipitation phase is $\sim$~400~K. Because we assume that water vapour is saturated 
everywhere in the atmosphere (except in the isothermal [155~K] stratosphere, whenever a stratosphere exists), the total 
water vapour content and thus the atmospheric CO$_2$ content can be calculated a posteriori. In our 1-D simulations, the 
total water vapour partial pressure ranges approximately between 0.1~Pa and 10~bar, and the CO$_2$ partial pressure between 3~mbar and 25~bar, in 
order to cover a wide range of possible post-impact atmospheric states.

From these simulations, we calculated the radiative disequilibrium at the top of the atmosphere (TOA) for various 
combinations of CO$_2$ and H$_2$O atmospheric contents, summarized in Figure~\ref{large_impacts_scatter_plot}. 
Our results are unequivocal. Whatever the amount of water vapour, CO$_2$ in the atmosphere, and whatever the initial 
post-impact temperature assumed, the atmosphere is always out of equilibrium, and there is no long-term 
greenhouse-induced self-maintained warm climate induced by the impact.

The only balanced solutions are found for very low surface temperatures (below 230~K), for CO$_2$ partial pressures on 
the order of a few bar, recovering the cold surface temperatures predicted by 1-D simulations of early Mars 
assuming CO$_2$/H$_2$O atmospheres \citep{Wordsworth:2010,Ramirez:2014,Turbet:2017jgr}. These solutions result 
from a subtle balance between the greenhouse effect of CO$_2$, CO$_2$ atmospheric condensation and CO$_2$ Rayleigh scattering.

In the upper part of the diagram, the outgoing longwave radiation (OLR) has an asymptotic behaviour whatever the CO$_2$ atmospheric pressure, 
whenever water vapour is a dominant species. At 1~bar of CO$_2$ and 2~bar of H$_2$O, we match the reference case described with 
the 3-D Global Climate Model in the previous section. The OLR and absorbed solar radiation (ASR) are 235 and 85~W/m$^2$, 
respectively, which gives a TOA radiative disequilibrium of -150W/m$^2$. The radiative disequilibrium is 30W/m$^2$ less than 
calculated in the 3-D simulation (see Fig.\ref{large_impacts_1D_mean_plots}B). Although we recover a very similar OLR, 
the ASR is much higher (+~30~W/m$^2$) in the 1-D simulation. The difference is likely due to the fact that the 3-D 
simulation accounts for the albedo of the thick cloud cover forming in the upper atmosphere, whereas clouds are not 
taken into account in the 1-D simulations presented in this subsection.

We confirm the result obtained in the previous section (with the 3-D GCM) that whenever water vapour 
becomes a dominant species, the OLR reaches asymptotic values (see Figure~\ref{large_impacts_scatter_plot}) that are very similar 
to those calculated by 1-D climate models for moist atmospheres in or near the runaway greenhouse \citep{Kopparapu:2013,Kopparapu:2014}. 
The fact that the OLR reaches -- in the water-rich limit -- an asymptotic value much larger than the solar 
flux possibly absorbed by early Mars is a strong argument against the \citet{Segura:2012} hypothesis, 
i.e. that stable runaway greenhouse states are stable on early Mars. 
Although two stable solutions (one cold, one warm) are indeed predicted by calculations 
assuming purely radiative H$_2$O-dominated atmospheres, the warm solution should be physically implausible 
\citep{Ingersoll:1969,Nakajima:1992,Goldblatt:2012} 
when convection and condensation processes are considered. First, this warm solution requires water vapour supersaturation 
levels that are extremely high (see \citealt{Goldblatt:2012}, Fig.~2), so high that they lie well above the maximum 
supersaturation limits (even imposed by homogeneous nucleation) of water vapour \citep{Pruppacher:1996}. Secondly, these 
purely radiative calculations neglect convective processes that control the thermal structure of the atmosphere. 
Whenever (i) convection processes are included and (ii) water vapour is limited by saturation, the bistability disappears 
and we recover the Nakajima limit \citep{Nakajima:1992,Goldblatt:2012}, i.e. the asymptotic behaviour of 
the OLR at the runaway greenhouse.




Whatever the initial reservoir of CO$_2$ and the amount of H$_2$O produced in response to the impact event, the duration of the warm period (i.e. for surface temperatures above the freezing point of water) is short. Figure~\ref{large_impacts_scatter_plot} provides estimates of the radiative disequilibrium at the TOA for many different combinations of CO$_2$ and H$_2$O reservoirs. These TOA radiative disequilibriums can be used to estimate the duration of the post-impact warm periods. For instance, for a 1~bar CO$_2$ atmosphere (similar to the 3-D reference simulation), the duration of the warm period (around 10-20~martian years) is very similar to that calculated with a 3D Global Climate Model. More generally, our calculations are in rather good agreement with the 1-D cloud-free climate calculations of \citet{Segura:2002} and \citet{Segura:2008}. 

The duration of the impact-induced warm period increases with increasing CO$_2$ atmospheric content (see Fig.~\ref{large_impacts_scatter_plot}) because as CO$_2$ atmospheric levels increase, the temperature - at a given atmospheric pressure - decreases. This results from the fact that the atmospheric temperature - at saturation - is governed by the partial pressure of water vapour (and not the total pressure). As a result, adding CO$_2$ cools the upper atmosphere, which drastically reduces the OLR. Fig.~\ref{large_impacts_scatter_plot} indicates that, for atmospheres made of 10+~bar of CO$_2$, the net TOA radiative budget could be reduced by a factor of ten and the duration of the impact-induced warm period could thus increase by a factor of ten, compared to the reference simulation presented in Section~\ref{impact_3D}. 
However, such high CO$_2$ atmospheric contents are unlikely \citep{Forget:2013,Kite:2019}.

\subsection{Results from cloudy numerical climate simulations}

We now include in our 1-D simulations the radiative effect of a cloud cover (as described in Section~\ref{large_impact_1Dmodel_method}). 
We use these simulations to explore how cloud microphysics (that depends in our 3-D Global Climate Model on the assumed number of 
cloud condensation nucleis [CCNs], and on the efficiency of the coagulation and sedimentation processes) could affect the results 
presented in the previous section.

\subsubsection{The radiative effect of water ice clouds: comparison with Ramirez et al. 2017}
\label{large_impact_ramirez_compare}

\begin{figure*}
\centering
\includegraphics[scale=0.37]{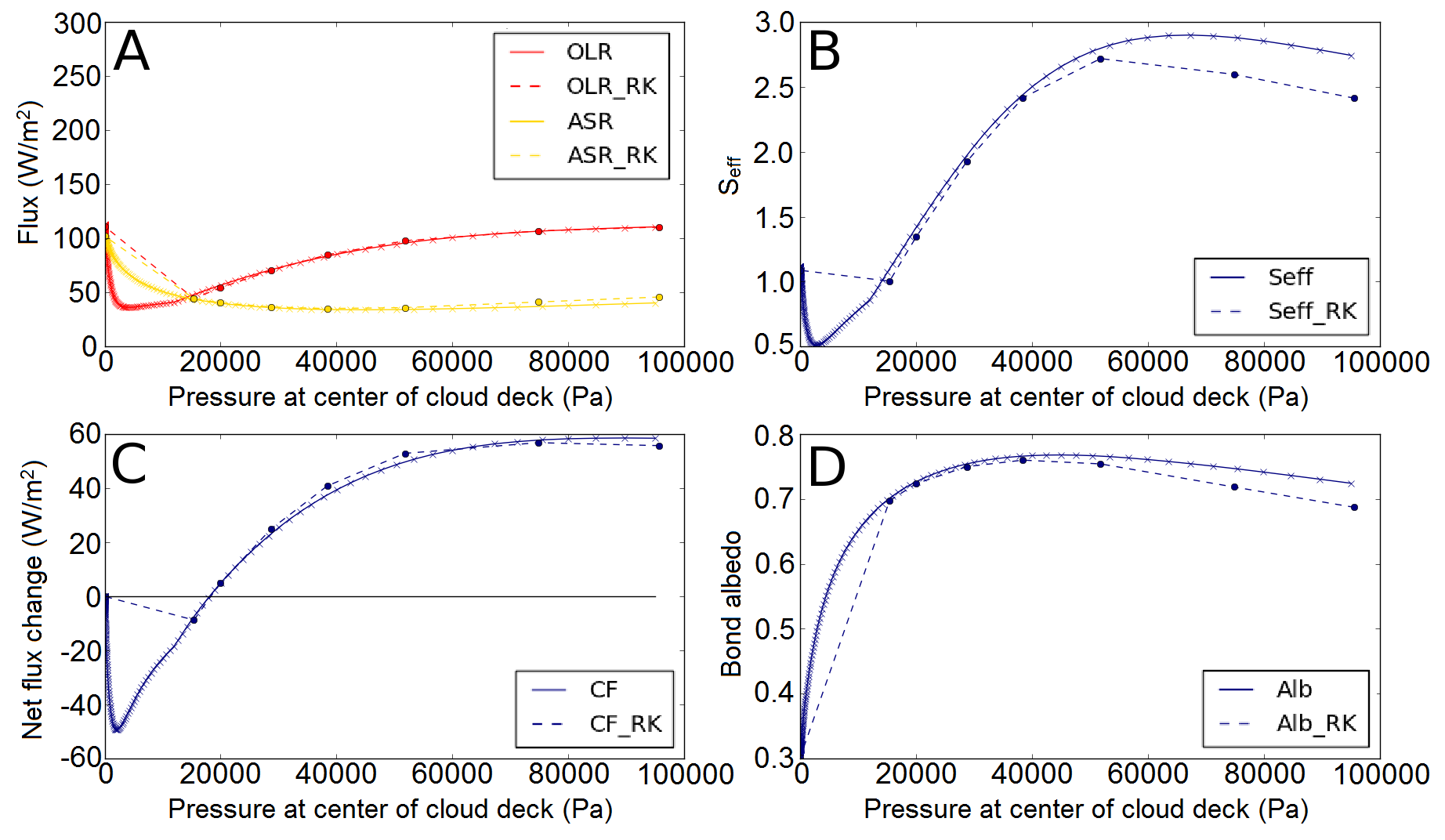}
\caption{Effect of a single cloud layer on the radiative budget of early Mars, as a function of the pressure at the center of the cloud deck. (A) Outgoing Longwave Radiation (OLR) and Absorbed Solar Radiation (ASR) ; (B) Effective solar flux S$_\text{eff}$ ; (C) Net flux change between cloudy and non-cloudy cases ; (D) Bond albedo. The dashed lines correspond to the results of \citet{Ramirez:2017} (presented in their Figure~2 and Table~1b). The solid lines correspond to the results obtained with our 1-D inverse climate model. We assumed here a 1~bar CO$_2$ dominated atmosphere (fully saturated in H$_2$O, except in the stratosphere [if present]). The cloud layer is assumed to be composed only of 10$\mu$m cloud particles. The surface temperature is assumed to be that of the freezing point of water, i.e. 273K. The solar flux is assumed to be that of present-day Mars. Note that the effective solar flux corresponds to the value of the solar flux (with respect to the Solar constant on Mars) required for the planet to be at the TOA radiative balance.}
\label{large_impacts_compare_rk17}%
\end{figure*}

We first compare the results of our 1-D inverse cloudy climate model in 'temperate' (surface temperature fixed to 273~K) 
conditions with the results of \citet{Ramirez:2017}. \citet{Ramirez:2017} used a very similar model to explore if 
cirrus clouds could have warmed the surface of early Mars above the melting point of water. Figure~\ref{large_impacts_compare_rk17} 
shows a comparison of the results of our model with theirs using similar assumptions. Whenever data points 
are available, the agreement between the two models is really good. We note a slight difference for the bond albedo of 
low altitude clouds (affecting subsequently the calculation of the effective flux S$_\text{eff}$ ; see Fig.~\ref{large_impacts_compare_rk17}BD) 
that is likely due to slight differences in the (visible) radiative properties of water ice particles.

\citet{Ramirez:2017} intentionally limited the maximum altitude of clouds explored in their simulations to the top of the (H$_2$O) 
moist convective region. Above, they claimed that production of water ice clouds should be unfeasible. 
This is why they did not provide any data point for pressures smaller than 0.15~bar. As a result, \citet{Ramirez:2017} did 
not capture the radiative effects of water ice clouds above the hygropause, although injections of water ice particles 
could be produced for example in response to extreme events such as bolide impact events as discussed in \citet{Urata:2013h2o}. 
Interestingly, this corresponds exactly to the altitude where the radiative effect of clouds is 
maximum (see Fig.~\ref{large_impacts_compare_rk17}C). This is not surprising because this is where the cloud 
temperature is minimal. For this reason, the linear interpolations proposed in Figures~2, 5, 8, 9 and 11 of \citet{Ramirez:2017} 
should be interpreted with great care by the readers.

Assuming that water ice clouds can exist above the hygropause, our model predicts that only a few tens of $\%$ of global cloud coverage should be sufficient to raise the surface of Mars above 273~K. This is illustrated in Figure~\ref{large_impacts_3Dplot_cloud_coverage}A that shows the minimum cloud coverage needed to reach a surface temperature of 273~K in the simulation (assuming a 1~km-thick cloud cover) depending on the surface pressure, cloud particle sizes and relative ice cloud water content (compared to the IWC [Ice Water Content]) defined in Section~\ref{large_impact_1Dmodel_method}. In theory, water ice cloud particles above the hygropause could thus warm early Mars above the melting point of water easily, even for relatively low (and thus reasonable) cloud coverage. 

However, we do not want to give the reader the impression that this provides a satisfactory scenario to 
warm early Mars. For clouds to be stable at such high altitudes, a strong mechanism must be at play to replenish 
the upper atmosphere in cloud particles that are expected to sedimentate rapidly \citep{Ramirez:2017}. The strong 
greenhouse warming of water ice clouds reported by \citet{Segura:2008} and \citet{Urata:2013h2o} is likely due to the 
fact that they neglected the sedimentation of ice particles. This hypothesis is supported by back of the envelope calculations 
of sedimentation rates of icy particles \citep{Ramirez:2017}, and confirmed by our 3-D Global Climate model simulations. 
As soon as the initial impact-induced water vapour reservoir is depleted, the production rate of cloud particles drops 
and upper atmosphere water ice clouds rapidly disappear because of gravitational sedimentation. Therefore, the lack of 
sedimentation is the most reasonable hypothesis to 
explain the difference between \citet{Segura:2008}, \citet{Urata:2013h2o} and our results. 
Until a plausible, long lasting source or mechanism of replenishment of upper atmospheric water ice clouds is evidenced, 
the strong water ice cloud greenhouse warming reported by \citet{Segura:2008} and \citet{Urata:2013h2o} is unlikely.

\subsubsection{Post-impact 1-D cloudy simulations}

\begin{figure*}
\centering
\includegraphics[scale=0.19]{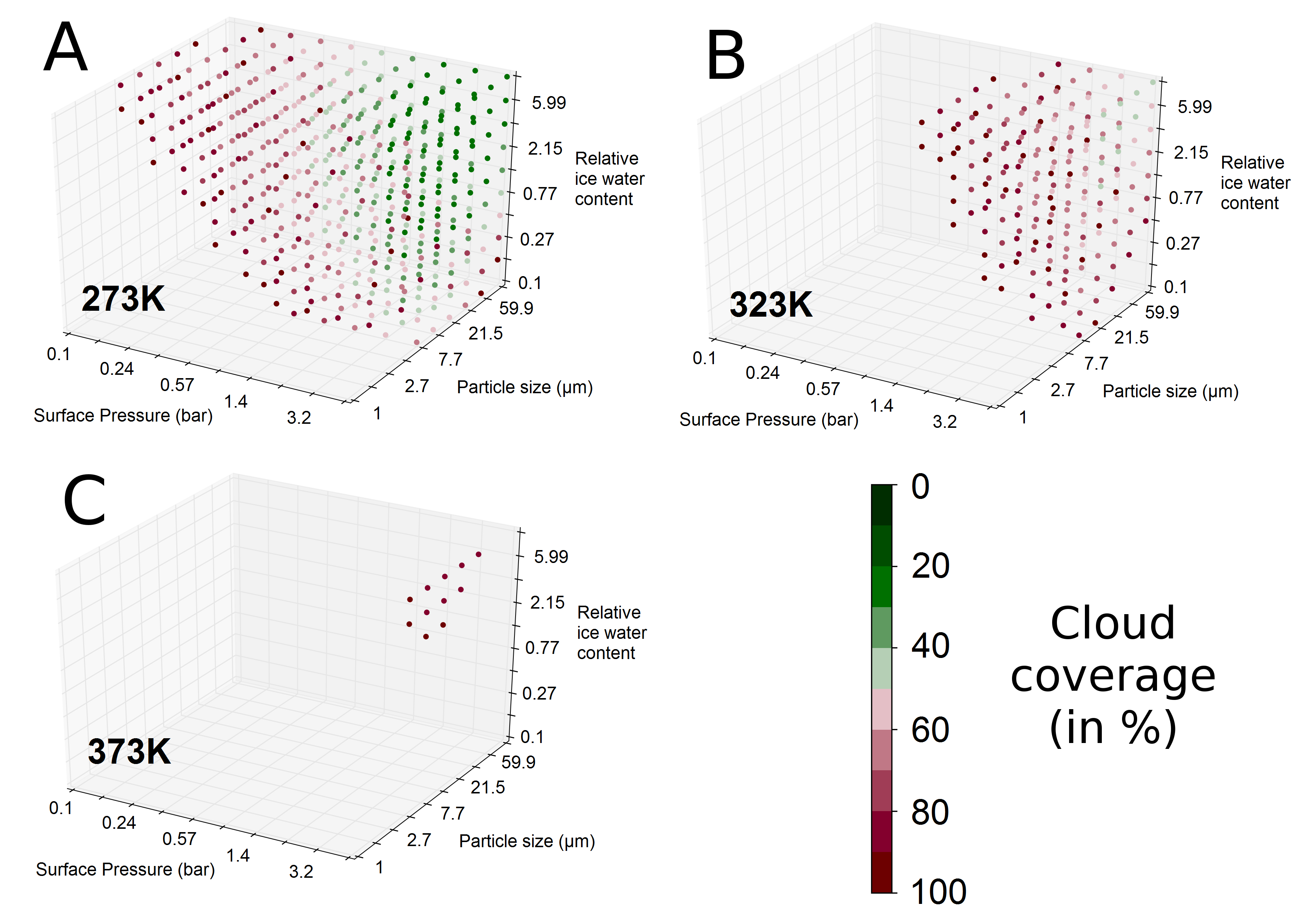}
\caption{3-D scatter plots of the minimal cloud coverage required to warm early Mars above the indicated 
post-impact surface temperatures (273, 323 and 373~K), for various CO$_2$ atmospheric pressure, 
relative ice water content (i.e. cloud thickness) and cloud particle size. Fractional cloud cover is modeled by averaging 
the radiative fluxes from cloudy and cloud-free 1-D simulations. No dot was plotted for sets of parameters unable to 
warm the planet above the indicated surface temperature even with a 100$\%$ cloud coverage. 
We provide in Figure~\ref{large_impacts_2Dplots_cloud_coverage} (Appendix) 2-D cross-sections of these 3-D scatter plots.}
\label{large_impacts_3Dplot_cloud_coverage}%
\end{figure*}

We showed in Section~\ref{large_impact_clouds} (using our 3-D climate simulations) that a thick, complete impact-induced cloud cover can be sustained in the upper atmosphere of early Mars for most of the duration of the main phase of precipitation following a large impact event. During this phase, surface temperatures are significantly higher than the value of 273~K explored above. Here we investigate how the cloud properties (size of particles, cloud thickness, etc.) can change the radiative effect of the impact-induced cloud cover and thus the net TOA radiative budget, during the post-impact main phase of precipitation.

Because we showed in the previous sections that (1) the thermal profiles calculated with our 1-D inverse climate model match very well those derived from 3-D Global Climate Model simulations during the main phase of precipitation, and that (2) the inclusion of a cloud layer in our 1-D model gives satisfactory results with regards to the existing literature \citep{Ramirez:2017} for a surface temperature of 273~K, we can now safely apply our 1-D cloudy numerical climate simulations to higher post-impact surface temperatures (T$>$~273~K). We use these simulations to investigate the radiative effect of the thick, complete cloud cover predicted by 3-D simulations.
 

Our findings are summarized in Figure~\ref{large_impacts_3Dplot_cloud_coverage}, that presents 
the minimal cloud coverage required to warm the surface of early Mars above the indicated surface 
temperatures (323 and 373~K). For the two surface temperatures considered, there is at least one 
combination of parameters (cloud altitude, cloud particles, cloud content and surface pressure) 
that can stably keep the surface temperature of early Mars above the indicated post-impact 
temperatures (323 and 373~K). However, although the constraints on (i) the possible total 
cloud coverage and (ii) the maximum altitude of the cloud layer are relaxed during the main 
post-impact phase of precipitation (as demonstrated with 3-D simulations in Section~\ref{large_impact_clouds}), 
the range of parameters that provide a positive radiative balance gets narrower as the surface temperature increases. 
In fact, the solutions that work - at high post-impact surface temperatures - are limited to very thick multi-bar 
CO$_2$-dominated atmospheres\footnote{This behaviour can be understood by looking back at Figure~\ref{large_impacts_scatter_plot} 
that shows that for post-impact cloud-free atmospheres, and for a fix water content/surface temperature, increasing the 
initial CO$_2$ atmospheric content decreases the TOA radiative disequilibrium. For example, for a surface temperature of 
373K, and a CO$_2$ initial surface pressure of 10~bar, the net TOA radiative disequilibrium is $\sim$~-40~W/m$^2$.} endowed 
with a very thick cloud cover located very high in the atmosphere, for cloud particles around 10~$\mu$m in size. This very 
restrictive set of conditions - although it could theoretically lead to a self-sustained impact-induced warm atmosphere - 
seems extremely difficult to attain.


\section{Exploration of the long-term effects of a very
large impactor on the interior of Mars with a 2-D Mantle Dynamics numerical model}
\label{stagyy_part}

Large impacts affect a planet like Mars on a broad range of timescales. 
Although its atmosphere reacts over a relatively short period, 
large collisions can modify the interior evolution
of the planet over millions to billions of years. 
Since the interior and
exterior of Mars interact, surface conditions are also affected by the
consequences of impacts on all timescales. In order to complete the
picture of environmental consequences of large impacts on Mars, we
included simulations of mantle dynamics effects of those events using the 
StagYY numerical code described in Section~\ref{large_impact_2D_mantle_dynamics_model_method}.

\begin{figure*}
\centering
\includegraphics[scale=0.42]{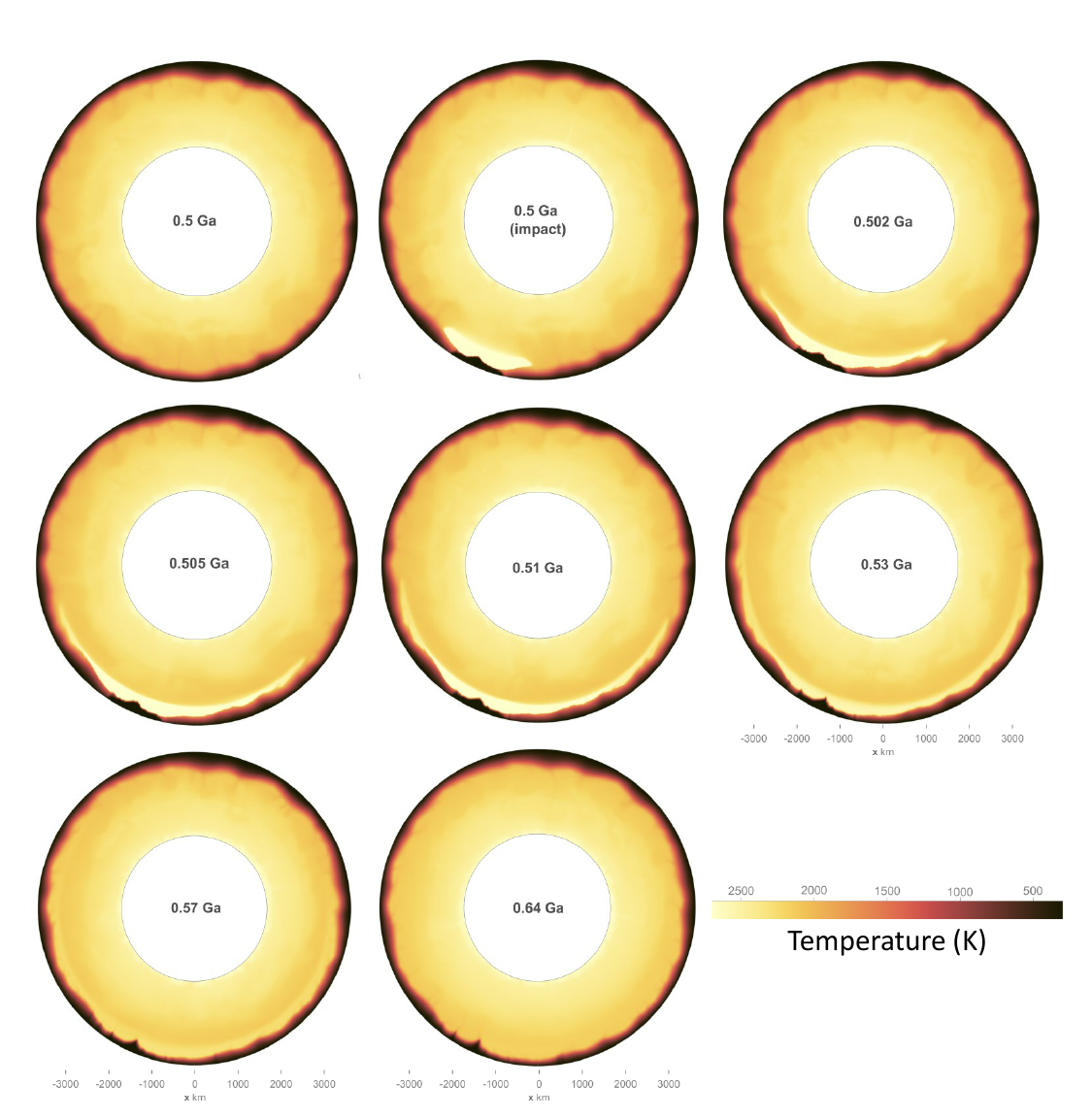}
\caption{Evolution of the martian mantle temperature following a 300~km diameter impactor 
(roughly corresponding to a crater diameter of 1500~km, using \citealt{Toon:2010} scaling relationship) hitting the surface of Mars (reference case).
The velocity of the impactor is set to set to 9~km$/$s for the reference case.}
\label{2D-annulus}%
\end{figure*}

We considered projectiles in the range of 200-500 km diameter (i.e., producing crater diameter of 1000-2500~km; \citealt{Toon:2010}); with our
reference case set at 300~km (crater diameter of 1500~km). Projectiles significantly smaller than the
reference case (D$_{\text{impactor}}$~$<$~200 km, i.e. D$_{\text{crater}}$~$<$~1000~km) only have a marginal effect on mantle
dynamics. Global changes in convection patterns are seen mainly if D$_{\text{impactor}}$~$>$~500~km 
(i.e. D$_{\text{crater}}$~$>$~2500~km), but such impacts are outside the range of observed events for the
present work (see \citealt{Gillmann:2016}).
Impactor velocities range from 5.1 to 20 km$/$s, with a reference case set to 9 km$/$s.

For high velocity cases, the impact-induced thermal anomaly
can reach temperatures up to 5000~K.
Figure~\ref{2D-annulus} illustrates that the anomaly decreases after the peak 
on a timescale of $\sim$~10$^7$ years. 
The maximum temperature of the anomaly is 3500~K and 2700~K after 1 and 10~million years, respectively.
After the emplacement of the buoyant temperature anomaly, 
a stage of thermal relaxation occurs, where the hot zone flattens under the 
surface of the planet and widens, recovering qualitatively the results of \citet{Gillmann:2016} on Venus and 
of \citet{Ruedas:2017,Ruedas:2018,Ruedas:2019} on Mars.
\begin{figure*}
\centering
\includegraphics[scale=0.19]{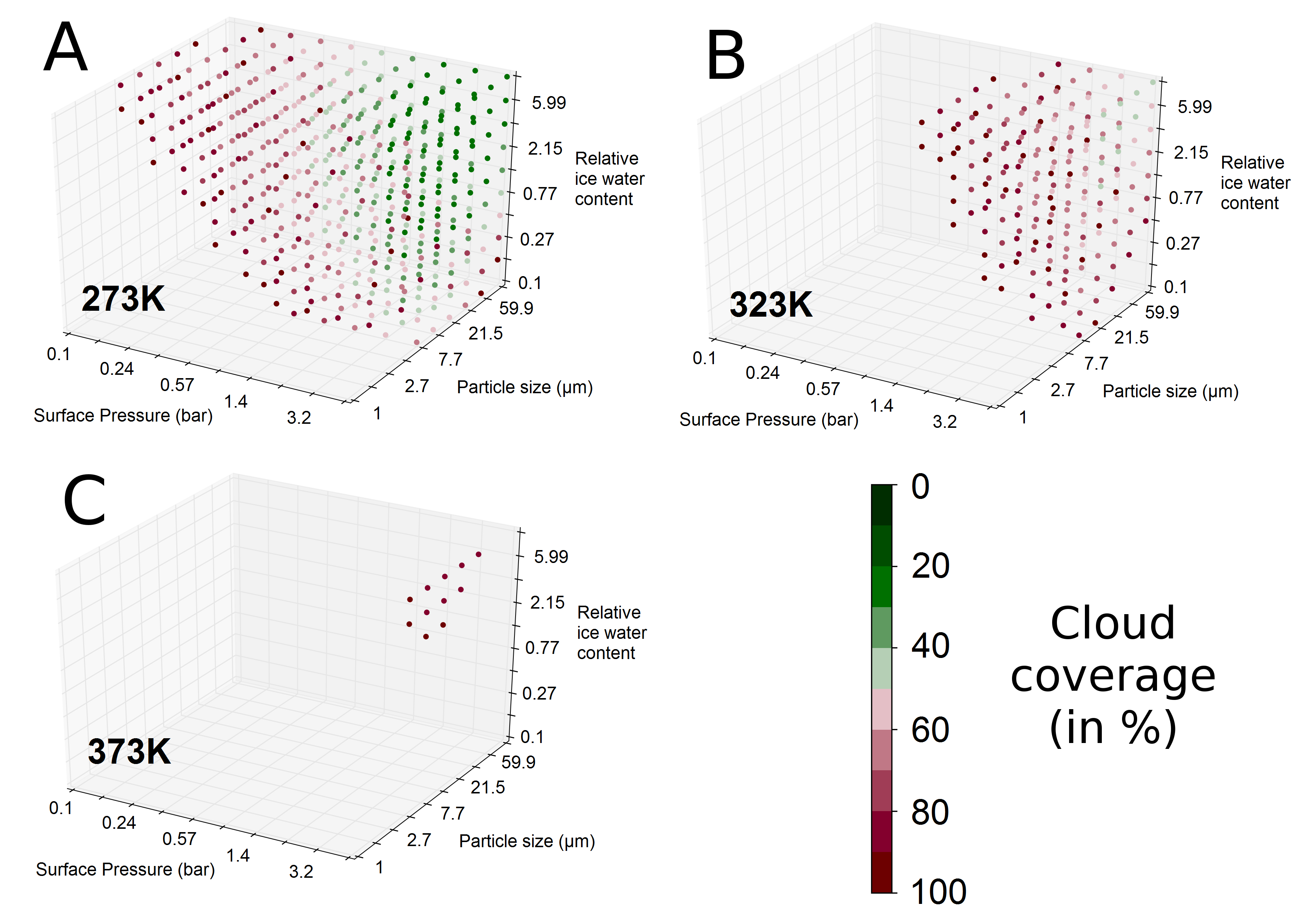}
\caption{3-D scatter plots of the minimal cloud coverage required to warm early Mars above the indicated post-impact surface temperatures (273, 323 and 373~K), for various CO$_2$ atmospheric pressure, relative ice water content (i.e. cloud thickness) and cloud particle size. Fractional cloud cover is modeled by averaging the radiative fluxes from cloudy and cloud-free 1-D simulations. No dot was plotted for sets of parameters unable to warm the planet above the indicated surface temperature even with a 100$\%$ cloud coverage.}
\label{large_impacts_3Dplot_cloud_coverage}%
\end{figure*}
Partial melting of the mantle occurs in the interior, where the thermal anomaly is maximum, i.e. near the impact location. 
The liquid melt-pond produced by the impact heating is in contact with the atmosphere and 
convects vigorously, as a consequence, it loses heat very efficiently and freezes quickly compared to 
the typical timescale of mantle convection \citep{Reese:2006,Solomatov:2015}. 
This produces the formation of a basaltic crust whose thickness is maximum at the location 
of the impact. Consequently, the crust is locally 
thickened, which reduces the efficiency of the heat transfer \citep{Lenardic:2004} near the impact location. 
This thick basaltic crust acts as an insulating layer in a comparable way to 
the effect of continents on Earth \citep{Coltice:2009}.

\begin{figure*}
\centering
\includegraphics[scale=0.4]{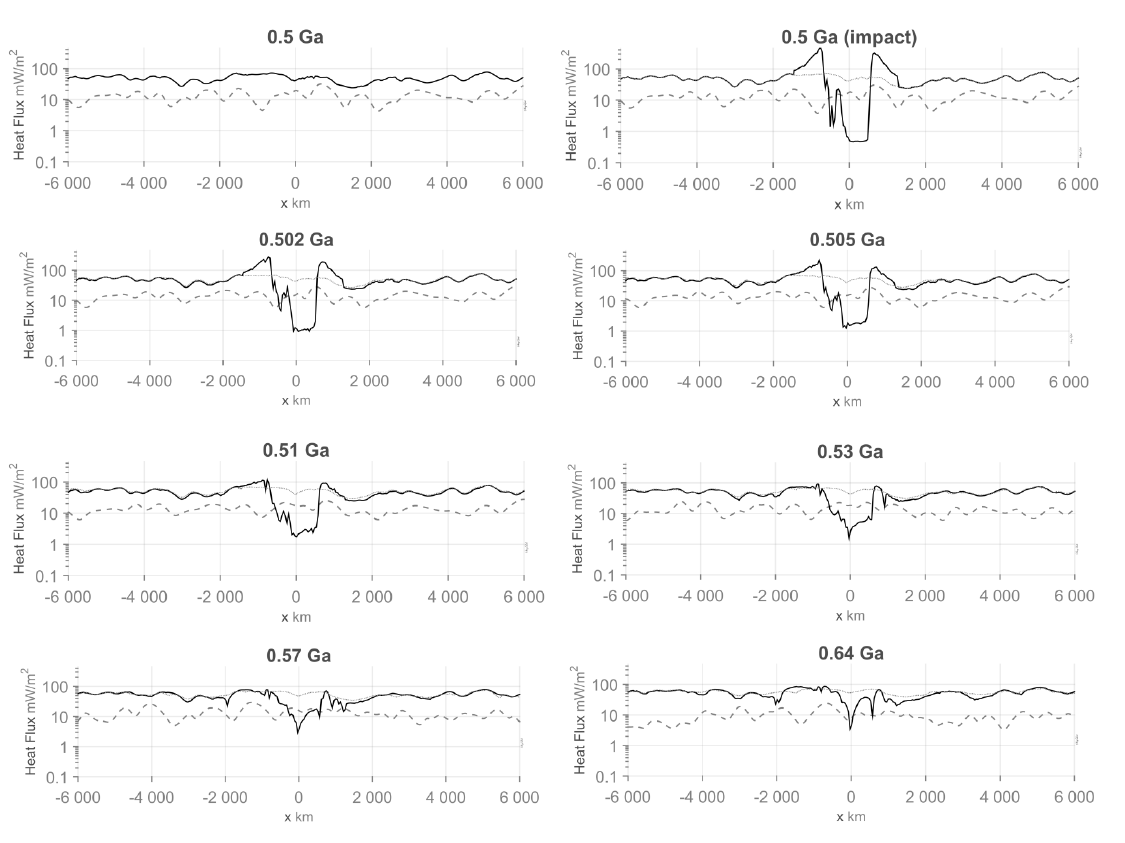}
\caption{Evolution of the near-surface internal heat flux (in mW~m$^{-2}$) following a 300~km diameter impactor 
(i.e. a crater diameter of 1500~km, using \citealt{Toon:2010} scaling relationship) 
hitting the surface of Mars (black line), as a function of the distance to the impact crater.
The grey, dotted line indicates the near-surface internal heat flux for the reference simulation without impact. 
The grey, dashed line indicates the internal heat flux at the Core Mantle boundary (CMB).
Note that the internal heat flux at the CMB is weakly affected by the impact.}
\label{2D-heat_flux}%
\end{figure*}

Figure~\ref{2D-heat_flux} shows the evolution of the heat flux across the surface following the large, reference bolide impact. 
At the impact location, the thermal anomaly is very strong but the 
surface internal heat flux is low\footnote{The internal heat flux could be very 
strong in the near subsurface for 10$^2$-10$^3$~years \citep{Segura:2002}, 
due to the residual of the near-surface impact-induced thermal anomaly as well as the hot ejecta deposited all over the planet. 
Another effect to consider is the potential impact melt sea that will collect on the floor of the basin, 
and that can take a longer time to cool to equilibrium \citep{Vaughan:2013,Vaughan:2014,Cassanelli:2016}. 
These effects are not captured by the StagYY code which explores the effects of the impact on much longer timescales.}
(much lower than before the impact) and can drop below 10~mW/m$^2$ 
for several tens of millions years (see Figure~\ref{2D-heat_flux}) because 
of the insulation produced by the basaltic crust. 
At the edge of the impact location, the thermal anomaly is still high, but low enough to limit the melting and thus the production of 
insulating basaltic crust\footnote{The impact-induced ejecta layer 
deposited over the martian surface could also insulate part of the interior. 
This effect is not accounted in our 2-D Mantle Dynamics simulation.}. 
As a result, the internal heat flux can reach several hundreds of mW/m$^2$ 
or $\sim$~10~times the ambient flux (see Figure~\ref{2D-heat_flux}) at the edge of the impact location for 
several millions of years.
Far from the impact location, the thermal anomaly is too low to significantly affect the internal heat flux.

In summary, although the surface internal heat flux can be very strong at 
the location of the impact for 10$^2$-10$^3$~years \citep{Segura:2002}, 
on the longer term this flux drastically drops below 10~mW/m$^2$ for tens of millions of years. 
Besides, the internal heat flux significantly increases up to several hundreds mW/m$^2$ 
(i.e. up to $\sim$~10~times the ambient flux) on the edges of the 
impact crater for millions of years. 
Although such internal heat fluxes are orders of magnitude too low to keep the martian surface above the melting point of water, 
they could have had an impact on the subsurface hydrothermal system and on the basal melting of ice deposits (that could have 
accumulated on the impact crater edges through atmospheric processes, described in Phase~IV; see Section~\ref{large_impact_chronology}). 
We leave this study for future work.

\section{Conclusions and Discussions}

We explored in this manuscript the environmental effects of the largest impact events recorded on 
Mars using a hierarchy of numerical models, ranging from a 1-D radiative-convective inverse model 
to a full 3-D Global Climate Model and a 2-D Mantle Dynamics model. 

Our results indicate that the duration of the impact-induced warm period (in the atmosphere and at the surface) is usually very short, 
because the radiative budget at the Top Of the Atmosphere (TOA) is in strong deficit. 
Whatever the initial CO$_2$ atmospheric content and whatever the size of the impactor, 
we show that the impact-induced stable runaway greenhouse state predicted by \citet{Segura:2012} is physically inconsistent. 

For an early martian atmosphere made of 1~bar of CO$_2$, a large impact would produce $\sim$2.6~m of 
surface precipitation per Earth year, until the reservoir of water vapour gets completely depleted. 
This is quantitatively similar to the results of \citet{Segura:2002}, obtained using a 1-D cloud free climate model. 
Surface evaporation of precipitation is weak (i.e. low amount of hydrological cycling of water) 
and as a result, the total amount of precipitation produced in response to the impact 
event can thus be well approximated by the total amount of water vapour initially vaporized and/or sublimed in the atmosphere 
(from the impactor, the impacted terrain and from the sublimation of permanent ice reservoirs heated by the hot ejecta layer). 

3-D simulations show that an optically thick, upper atmospheric cloud cover forms uniformly on the planet. Our 3-D simulations - 
taking into account simplified cloud microphysics - indicate that this cloud cover contributes to a net cooling of 
the surface of Mars, compared to cloud-free calculations. Although strong precipitation is generated in this cloud cover, 
most of it is re-evaporated in the atmosphere while falling. Instead, surface precipitation patterns are governed by lower 
atmospheric large scale circulation. 
In our 3-D simulations, precipitation (rainfall) is maximum near the equator, at the 
ascending branch of the Hadley cell, and are minimum near the tropics, at the descending branch. Although the cumulative 
amount of precipitation is rather high everywhere on the planet (at least 20~m of cumulative precipitation), we find that 
the main region of valley networks formation on Mars (located around -30$^\circ$S in the pre-TPW topography, see Fig~1 
in \citealt{Bouley:2016}) coincides with a minimum of cumulative precipitation in our 3-D simulations (see Fig~\ref{large_impacts_rain_map}).
We confirm the results of \citet{Segura:2008} and \citet{Urata:2013h2o} that water ice clouds could in theory 
significantly extend the duration of the post-impact warm period, and for significantly lower cloud coverage than predicted in 
\citet{Ramirez:2017}. However, as soon as the main bulk of the water 
vapour reservoir has condensed on the surface, the production rate of upper atmospheric cloud particles drops, 
and high altitude water ice clouds dissipate because of gravitational sedimentation. This prevents the water ice 
cloud greenhouse warming mechanism from working.
We also find that the duration of the impact-induced warm period increases with increasing CO$_2$ atmospheric content. For instance, the duration can be increased by a factor 10 for a 10~bar CO$_2$ atmosphere (with respect to a 1~bar CO$_2$ atmosphere). 
However, such high CO$_2$ contents are unlikely \citep{Forget:2013,Kite:2019}.

In summary, the environmental effects of the largest impact events recorded on Mars are characterized by (i) 
a short impact-induced warm period, (ii) a low amount of hydrological cycling of water (because the
evaporation of precipitation that reached the ground is extremely limited), 
(iii) precipitation patterns that are uncorrelated with the observed regions of valley networks formation, 
and (iv) deluge-style precipitation. 

On the long-term, we find that large bolide impacts can produce a strong thermal anomaly in the mantle of Mars that can survive 
and propagate for tens of millions of years. This thermal anomaly could raise the near-surface internal heat flux up to several hundreds 
of mW/m$^2$ (i.e. up to $\sim$~10~times the ambient flux) for several millions years at the edges of the impact crater. 
However, such internal heat flux is largely insufficient to keep the martian surface above the melting point of water.

\medskip

In addition to the 
poor temporal correlation between the formation of the largest basins and valley networks \citep{Fassett:2011},
these arguments indicate that the largest 
impact events are unlikely to be the direct cause of formation of 
the Late Noachian valley networks.
Our numerical results support instead the prediction of \citet{Palumbo:2018impact} that such 
deluge-style rainfall could have caused 
crater degradation of large craters, erased small craters, and formed smooth plains, 
potentially erasing much of the previously visible morphological surface history.
Such hot rainfalls may have also led to the formation of aqueous alteration 
products on Noachian-aged terrains \citep{Palumbo:2018impact}, which is consistent with the timing of clays that were expected to have 
formed in situ (e.g. \citealt{Carter:2015}).

\section{Acknowledgments}

This work was granted access to the HPC resources of the institute for computing
and data sciences (ISCD) at Sorbonne Universite. This work benefited from the IPSL ciclad-ng facility.
We are grateful for the computing resources on OCCIGEN (CINES, French National HPC). 
C. Gillmann was supported by BELSPO’s Planet TOPERS IAP programme and FNRS’s ET-HOME EOS programme. 
We thank P. Tackley for the use of his mantle dynamics code StagYY and
for his advice. The geodynamic diagnostics and scientific visualisation
software StagLab \citep{Crameri:2017,Crameri:2018} is used in this study. 
M.T. thanks Fuxing Wang, Frederique Cheruy, Jim Kasting and the LMD Planeto team for useful discussions related to this work.
This project has received funding from the European Union’s Horizon 2020 research and 
innovation program under the Marie Sklodowska-Curie Grant Agreement No. 832738/ESCAPE. 
M.T.  would  like  to  thank  the  Gruber  Foundation  for  its  generous  support  to  this  research.

\bibliography{Thesis}
\bibliographystyle{apalike}

\appendix

\section{Additional Figure}

\begin{figure*}
\centering
\includegraphics[scale=0.13]{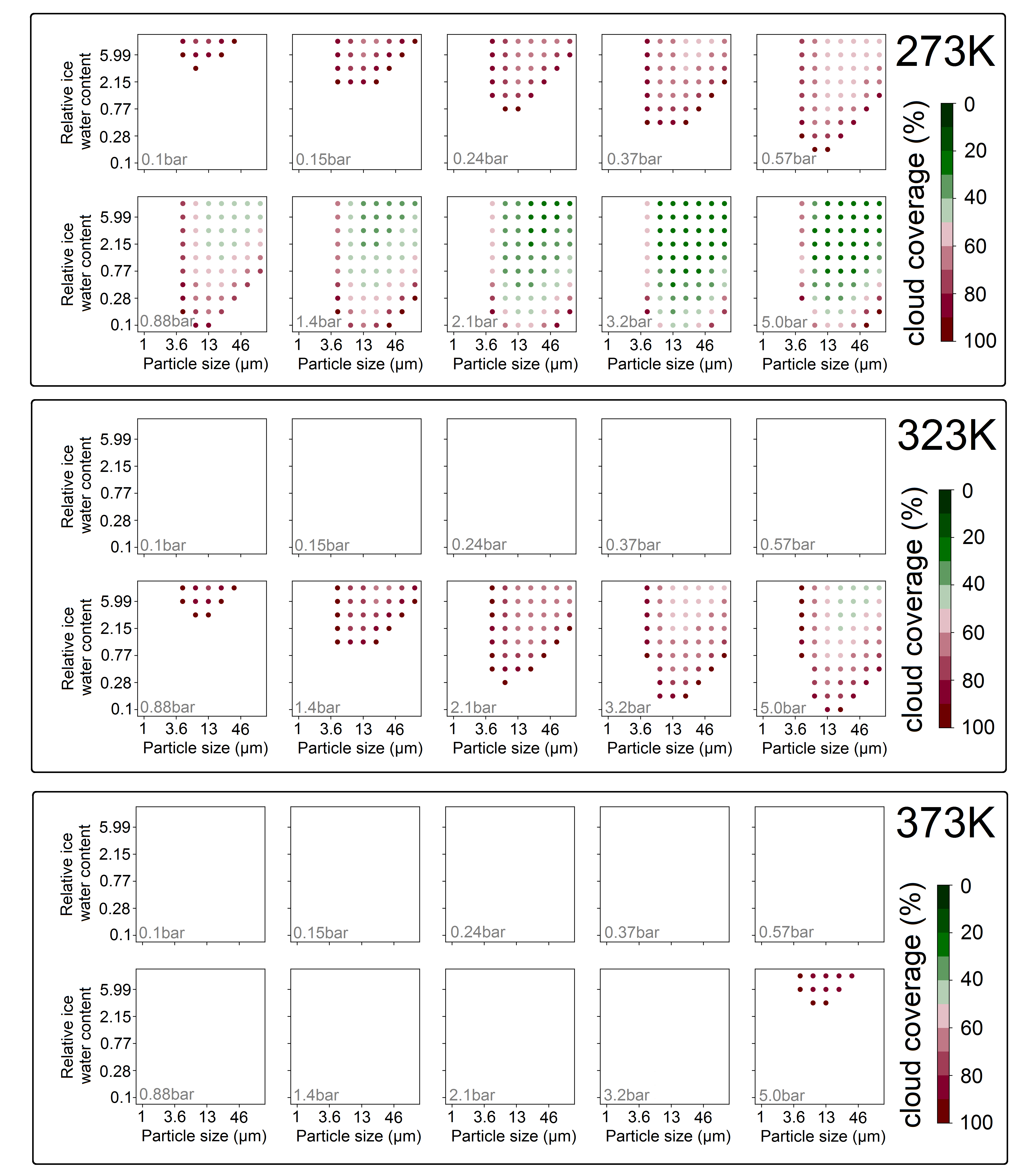}
\caption{2-D scatter plots of the minimal cloud coverage required to warm early Mars above the 
indicated post-impact surface temperatures (273, 323 and 373~K), for various CO$_2$ 
atmospheric pressure, relative ice water content (i.e. cloud thickness) and cloud particle size. 
These 2-D scatter plots are cross-sections of 3-D scatter plots presented in Figure~\ref{large_impacts_3Dplot_cloud_coverage}, 
for 10 distinct CO$_2$ surface pressures.}
\label{large_impacts_2Dplots_cloud_coverage}%
\end{figure*}

\end{document}